\documentclass[useAMS,usenatbib]{mn2e}
\usepackage{epsfig,graphics,graphicx,bm,amssymb}

\begin{document}

\title[X-ray emitting young stars associated with Sh~2-296]{Spectroscopic characterization of X-ray emitting young stars associated with the Sh~2-296 nebula\thanks{Based on observations obtained at the Gemini Observatory, which is operated by the 
Association of Universities for Research in Astronomy, Inc., under a cooperative agreement 
with the NSF on behalf of the Gemini partnership: the National Science Foundation 
(United States), the National Research Council (Canada), CONICYT (Chile), the Australian 
Research Council (Australia), Minist\'{e}rio da Ci\^{e}ncia, Tecnologia e Inova\c{c}\~{a}o 
(Brazil) and Ministerio de Ciencia, Tecnolog\'{i}a e Innovaci\'{o}n Productiva (Argentina).}}
\author[Fernandes et al.]{B. Fernandes$^{1}$\thanks{E-mail:
beatriz.fernandes@iag.usp.br}, J. Gregorio-Hetem$^{1}$, T. Montmerle$^{2}$ and
G. Rojas$^3$\\
$^{1}$Universidade de S\~ao Paulo, IAG, Rua do Mat\~ao 1226, 05508-900 S\~ao Paulo, Brazil,
\\
$^{2}$Institut  d'Astrophysique de Paris, France;
$^{3}$Universidade Federal de S\~ao Carlos, SP, Brazil}

\date{ }

\pagerange{\pageref{firstpage}--\pageref{lastpage}} \pubyear{2014}

\maketitle

\begin{abstract}
We studied a sample of stars associated with the Sh~2-296 nebula, part of the reflection nebulae complex in the region of Canis Major (CMa R1). Our sample corresponds to optical counterparts of X-ray sources detected from observations with the  {\it XMM-Newton}  satellite, which revealed dozens of possible low-mass young stars not yet known in this region.

A sample of 58 young star candidates were selected based on optical spectral features, mainly H$\alpha$ and lithium lines, observed with multi-objects spectroscopy performed by the {\it Gemini South} telescope. Among the candidates, we find 41 confirmed T Tauri and 15 very likely young stars. Based on the H$\alpha$ emission, the T Tauri stars were distinguished between classical (17\%) and weak-lined (83\%), but no significant difference was found in the age and mass distribution of these two classes.

The characterization of the sample was complemented by near- and mid-infrared data, providing an estimate of ages and masses from the comparison with pre-main-sequence evolutionary models. While half of the young stars have an age of 1-2 Myrs or less, only a small fraction ($\sim 25\%$) shows evidence of IR excess revealing the presence of circumstellar discs. This low fraction is quite rare compared to most young star-forming regions, suggesting that some external factor has accelerated the disc dissipation.

\end{abstract}

\begin{keywords}
 ISM: individual (Sh~2-296) -- stars: pre-main-sequence 
\end{keywords}

\section{Introduction}

Sh~2-296 is a bright rimmed nebula associated with the Canis Major R1 Galactic star-forming region.
The source of the shock front that originated the arc-shaped nebula and triggered star formation in this 
region is still uncertain (Elmegreen \& Lada 1977, Herbst \& Assousa 1977, Reynolds \& Ogden 1978,
Blitz 1980, Pyatunina \& Taraskin 1986).

To help elucidate the star formation history of Canis Major R1, 
\cite{GregorioHetem2009} conducted a wide-field {\sl ROSAT}
study around Sh~2-296 revealing that the star formation activity has been going on for more than 10 Myr.
This extended period of star formation is indicated by the existence of at least two groups 
of young stars  with different ages, both including several tens of low-mass stars. This result
was based on the discovery and characterization of a previously unknown cluster that 
is close to GU CMa and $\sim 30'$ away from the well-known young stellar groups close to Z CMa. 
The near-infrared (near-IR) characterization also suggested the presence in both
clusters of a small, but significant fraction of young ($<$ 5 Myr) and older ($>$ 10 Myr) stars. 

The scenario of sequential star formation has been explored in several stellar groups and
clouds. For instance, the young cluster NGC~6530, which shows the coexistence of  stars with
ages between  1-2  Myrs  and 6-7 Myrs, and masses between 0.4 and 4 M$_{\odot}$. Based on 
low-resolution spectra obtained with VIMOS/VLT, \cite{Prisinzano2012} derived the stellar
parameters for 78 members of NGC~6530 and found the presence of two distinct generations of 
 young stellar objects (YSOs) showing different spatial distribution.
   
\cite{Cusano2011} conducted a spectroscopic and photometric study aiming to characterize 
a sample of 23 pre-main-sequence (PMS) stars in Sh~2-284, an HII region that contains several 
young clusters.
The estimate of effective temperature, mass and ages of the sample indicates that triggered
star formation is occurring in this region, where a large fraction of the YSOs have preserved their
disc/envelopes.

In the Lupus molecular clouds \cite{Mortier2011}  selected  92 candidates
 among the objects showing near-IR excess as discovered by the {\it Spitzer} Legacy Program 
``{\it From Molecular Cores to Planet-Formings}". 
The effective temperature and luminosity
were derived for 54 objects, which are  mostly M-type stars, and 10\% are K-type stars. 
Depending on the adopted evolutionary  model, the mean population age is found to be between 
3.6 and 4.4 Myr and the mean mass is found to be $\sim$0.3M$_{\odot}$. The distribution of
spectral types of the  Lupus stellar population is similar to that in Chamaeleon~I   (Luhman 2007) and 
IC~348  (Luhman et al. 2003). The H$\alpha$ line was used to distinguish between classical and weak-line T Tauri
stars (respectively CTT and WTT) and revealed that  25 of the objects are accreting T Tauri stars. 

An impressive large sample of YSOs associated with the L~1641 cloud was recently characterized by \cite{Fang2013}  using multi-wavelength data ({\it Spitzer, WISE, 2MASS}, 
and {\it XMM-Newton}) to derive spectral types, extinction values, masses, ages,  accretion rates, and 
a disc fraction of the L~1641 members. \cite{Fang2013}  verified that the field stars in the
L~1641 region  show a bimodal distribution with peaks around spectral  types G0 and early-M.
On the other hand, the YSOs peak around early-M and most of them lie in between the 0.1 and 
3 Myr isochrones in the H-R diagram. The lithium depletion trend in young 
PMS stars was also used to estimate the stellar ages, by evaluating W(Li), 
the  equivalent  width of the Li  $\lambda$6708 absorption line, as a function of spectral 
type for the YSOs in the L~1641 sample. From this study, Fang et al. noted that WTT
with similar spectral types show a large scatter of  W(Li), probably due to an age spread 
of these objects.

Seeking to compare the discless with disc-bearing members of the Serpens Molecular Cloud, \cite{Oliveira2013b} obtained optical spectroscopy, complemented with photometric data, for a large number of candidates discovered with {\it XMM-Newton}.
The young nature of 19 discless stars was confirmed by determining their stellar temperatures and luminosities in order to estimate mass and age of these new Serpens members.
From a comparison with their previous results on Serpens \cite{Oliveira2013a} they found for both samples, discless and disc-bearing stars, a similar distribution of mass and age. \cite{Oliveira2013b} suggest that this similarity can bring new insight on  disc evolution studies, if the same kind of result is also confirmed for larger samples of discless candidates.

A detailed study of a young stellar population and its respective 
star-forming region, such as the examples mentioned above,  is unique because it requires a complete data set of 
sensitive spectroscopic observations. In the present work we selected a sample of 83 stars distributed in an area of $\sim$ 12 pc$^2$
(0.04 sq. deg.) near the edge of  Sh~2-296. Seeking to confirm the youth of the optical counterparts of X-ray sources detected in this star forming region, we focus our analysis on the spectral features
that are typically found in PMS stars.

We present the results of a  spectroscopic and photometric
follow-up of Sh~2-296 using {\it Gemini+GMOS} observations and photometric archival data to achieve a much improved characterization of 
the young stellar population associated with the nebula, aiming to add crucial information 
to understanding the complex scenario related to this region and many others showing similar 
star formation episodes. 

The paper is organized as follows. In Sections 2 and 3 we present our
observations, datasets and describe the data reduction technique. In Section 4 we discuss the identification and classification of the young stars in the sample and present the  spectral types and physical parameters determined by us. In Section 5 we analyse the photometric characteristics of the sample. In Section 6 we confirm the classification of some individual objects and finally our concluding remarks are presented in Section 7.


\begin{figure*}
\centering
\includegraphics[angle=270,width=0.6\textwidth]{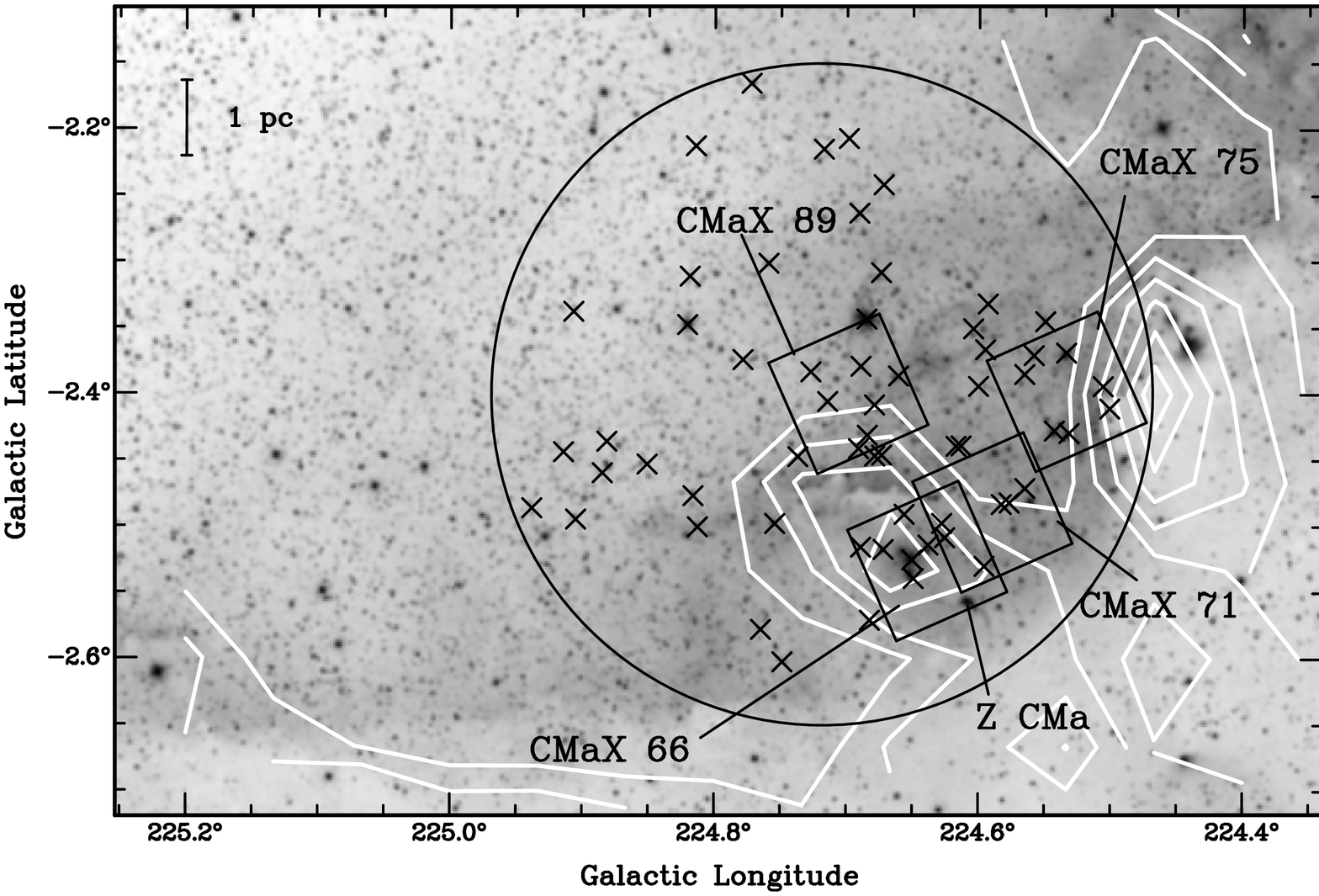}\\
\includegraphics[height=7.5cm,angle=270]{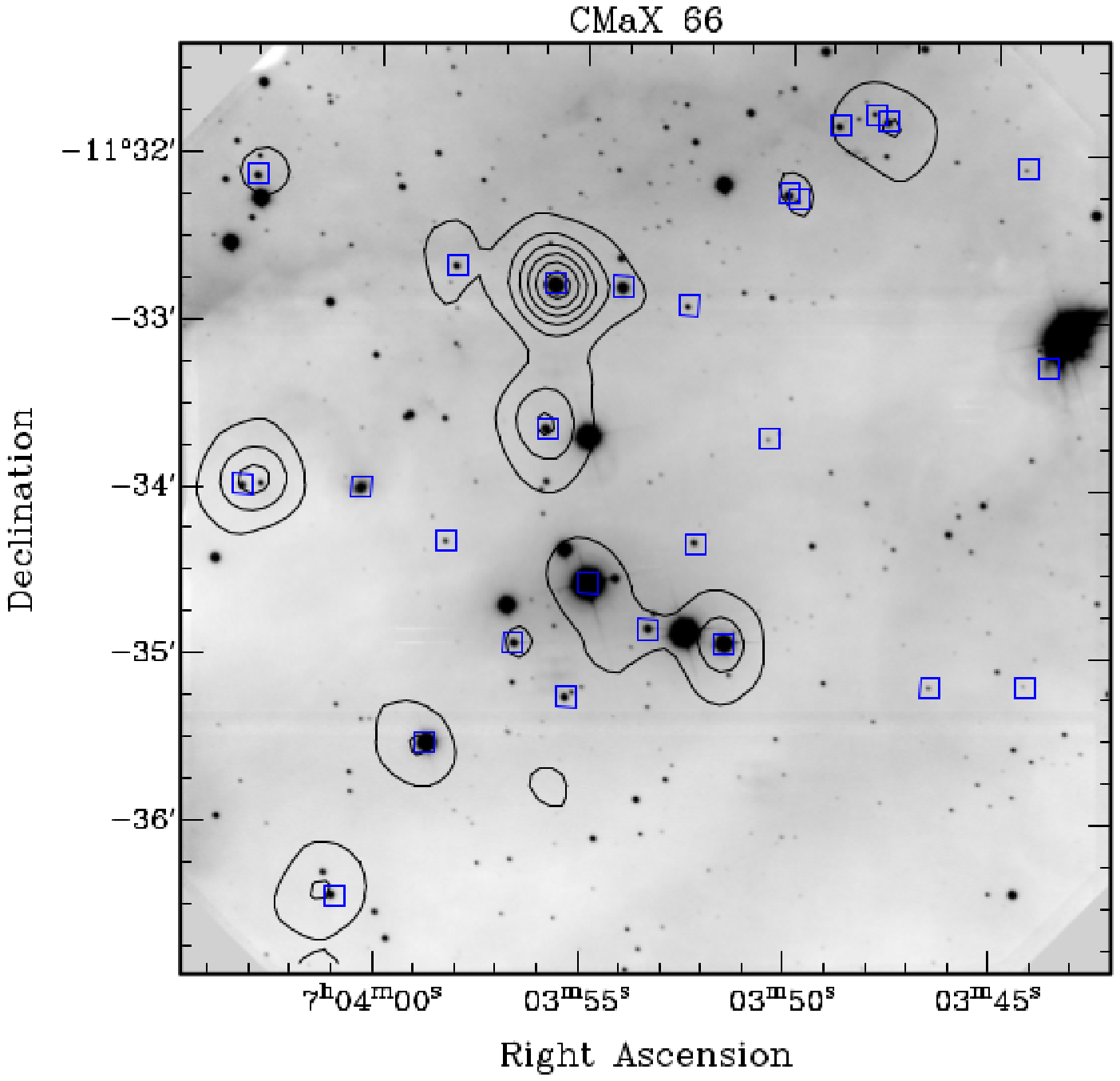}
\includegraphics[height=7.5cm,angle=270]{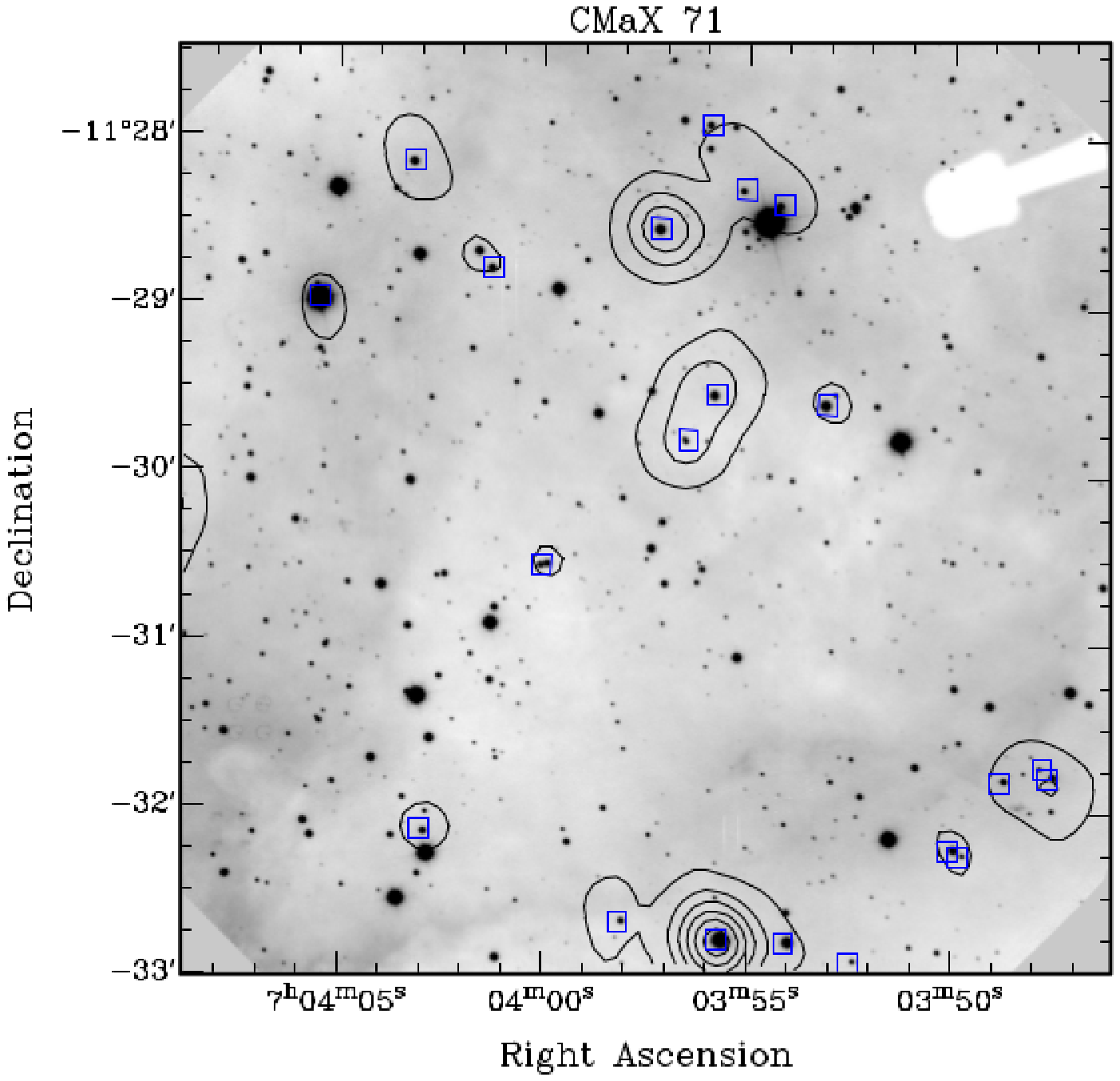}\\
\includegraphics[height=7.5cm,angle=270]{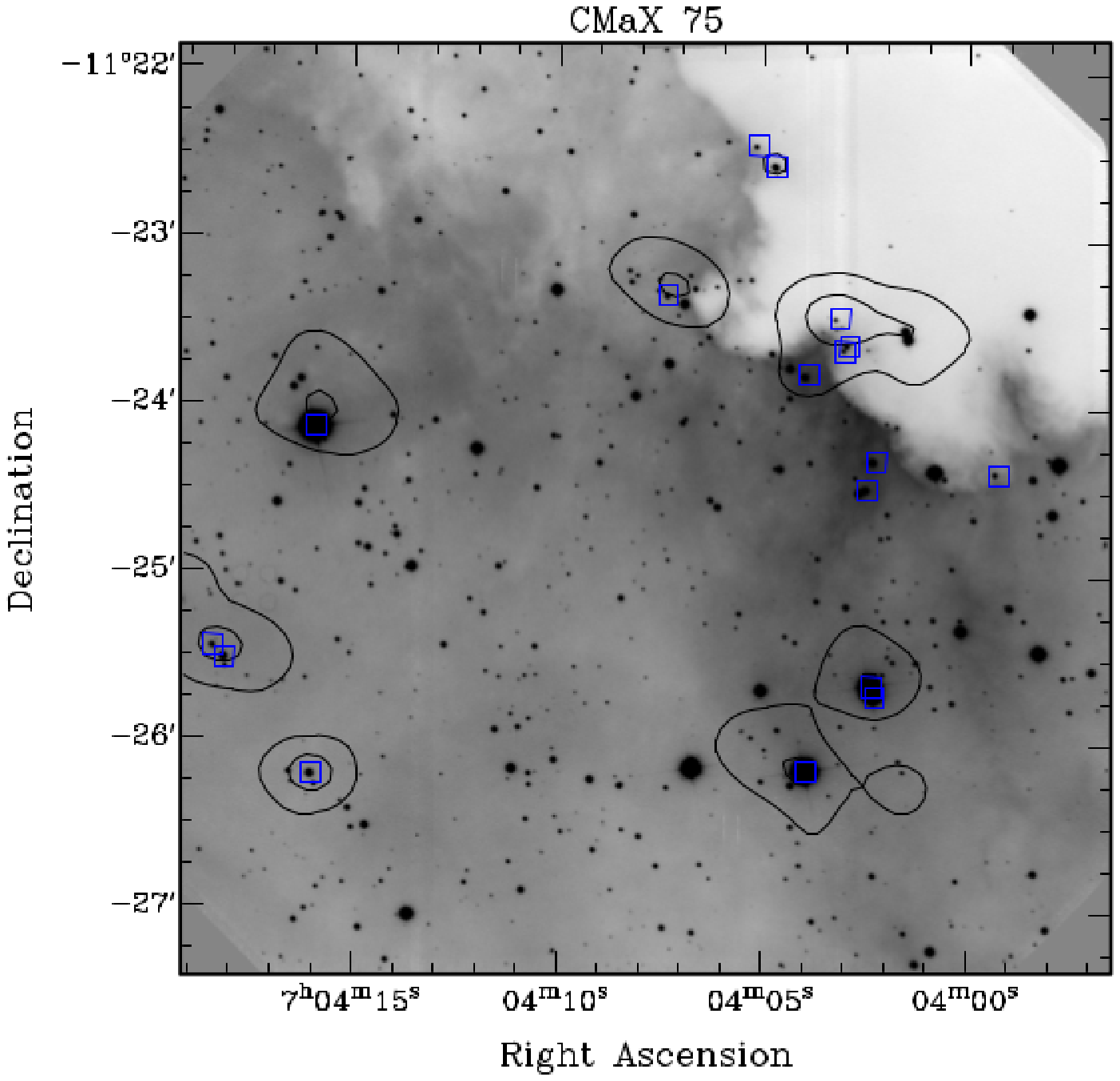}
\includegraphics[height=7.5cm,angle=270]{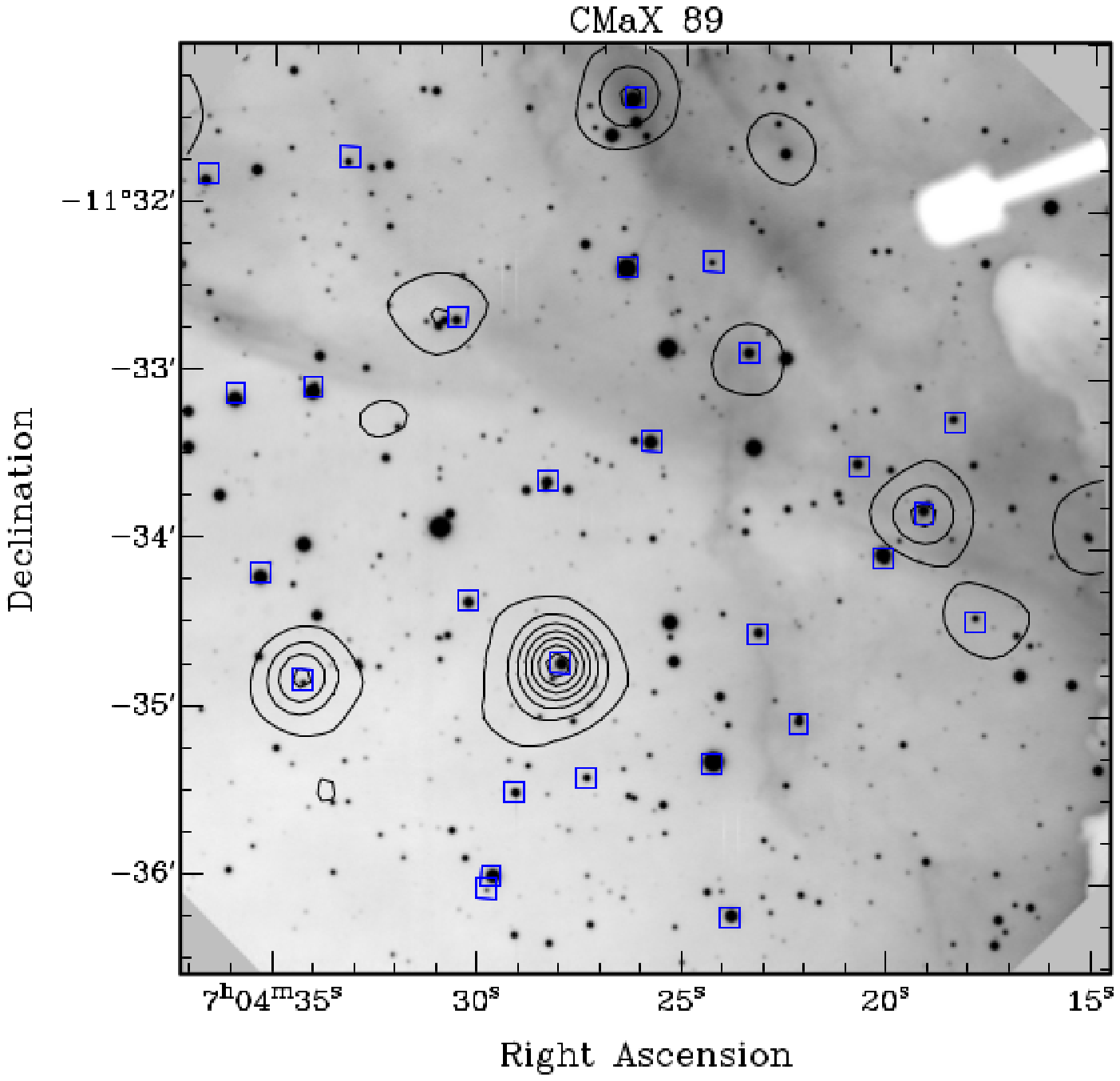}

\caption{{\it Top:} Optical (DSS) image of Sh~2-296 over-imposed by CO contours, {\it GMOS} FoVs (small squares), and X-ray sources (crosses) observed in the {\it XMM} FoV (large circle). {\it Middle and Bottom:}
R band image acquired with {\it GMOS} for the four regions covered by our observations superimposed by X-ray contours of sources detected by {\it XMM} (in black). Open squares indicate the sources for which  spectra was acquired.}
\label{fieldtot}
\end{figure*}


\section{Sample Selection based on {\it XMM-Newton} observations}

Aiming to improve the young stellar population census in the CMa~R1 region, our team performed a {\it XMM-Newton} EPIC observation covering young stellar clusters embedded in the Sh~2-296 nebula \citep{Rojas2006}. The observation was heavily affected by flaring background during $\sim$ 90\% of the allocated time, preventing us to use PN data. Nevertheless, a total of 61 X-ray sources were detected in this observation. 

Additionally, we retrieved {\it Chandra ACIS-S} archival data (ObsIDs 3751 and 10845) around the star Z CMa. These additional data accounted for 
140 X-ray sources, 103 of them not detected  by the {\it XMM-Newton} observation.

Details on these X-ray observations are given in Table 1, among with previous {\it ROSAT} observations. The sources identified by \cite{Rojas2006} are located near the star Z~CMa and other stellar groups like  the young cluster VdB~92 \citep{vandenBergh1966} and the YSOs associated with the cloud BRC~27 \citep{Sugitani1991}, which are found at the outer edge of Sh~2-296. Soares \& Bica (2002, 2003) estimated ages of 5-7 Myr for VdB~92, and 1.5 Myr for stars in  BRC~27, based on near-IR data and colour-magnitude diagrams.

More recently, BRC~27 was studied by \cite{Rebull2013} that used mid-IR data from {\it Spitzer} telescope to search for YSOs associated with bright-rimmed clouds. Their study identified new YSO candidates, revealed by the {\it Spitzer} data, and also included known YSOs from literature based on cross-identifications with several works covering the CMa R1 region (Wiramihardja et al. 1986, Shevchenko et al. 1999, Ogura et al. 2002, Chauhan et al. 2009). More details on the stellar identifications and the conditions for star formation in CMa R1 are given by \cite{Gregorio2008} in the Handbook of Star Forming Regions (ed. Bo Reipurth).
In the Sh~2-296 region, only a few X-ray sources were previously detected with {\it ROSAT}, that are named ``CMaX" by \cite{GregorioHetem2009}. In particular, CMaX~74 and  CMaX~75, which  are associated with the cloud BRC~27, were also studied by \cite{Rebull2013}. 

In spite of the presence of several YSO candidates around Sh~2-296, which were identified on the basis of near-IR data, there is a lack in the literature of a multi-spectral analysis for a large sample of objects in this region. This kind of study is required to reveal similarities and differences on age and spatial distribution of the young stars.

In total, the {\it XMM-Newton} observations and the {\it Chandra} archival data revealed 164 X-ray sources, spatially distributed in stellar groups previously unresolved by 
{\it ROSAT}. The correlation of near-IR counterparts with these X-ray sources indicates that they are probably young stars. Aiming to confirm the nature of some of  these 
candidates, we performed optical spectroscopy in four $5' \times 5'$ fields containing  37 of the X-ray sources identified by \cite{Rojas2006}.  In order to cover as
 many sources as possible, the fields for spectroscopy were centered on the main concentrations found in the spatial distribution of sources. A mean number of 9 X-ray sources is found in the chosen fields, which cover  11 to 28 observed stars each. In total, the sample has 83 objects: 45 coincide with
X-ray sources (8 of them in multiple systems), while the other 38 stars were observed with lower priority, since they are not associated with X-ray sources. These
additional candidates were included in the sample with the sole purpose to complete the observable field. Table 2 summarizes the number of observed stars in each 
field that are labeled according the presence of  a representative ``CMaX" source previously identified by \cite{GregorioHetem2009}.

Firgure \ref{fieldtot} (top panel) displays the FoVs of the {\it XMM} and {\it GMOS} observations compared to optical
DSS image and CO map \citep{Mizuno2004} showing the gas distribution, which coincides
with the dense regions around the nebula border.


\begin{table} 
\caption{X-ray observations in the CMaR1 region.}
\begin{center}
{\scriptsize
\begin{tabular}{lccccc}
\hline
& Obs. & Exp & Tot. & Net & area \\
Detector & ID & (ks) & src$^e$ & src$^e$ & (sq. deg.) \\
\hline
Chandra/ACIS & 10845$^a$ & 40 & 86 & 86 & 0.08 \\
Chandra/ACIS & 3751$^b$ & 38 & 63 & 17 & 0.08 \\
XMM/EPIC& 0201390201$^c$ & 3.1 & 61 & 37 & 0.196 \\
ROSAT/PSPC & RP201277$^d$ & 4.6 & 42 & 22 & 4 \\
ROSAT/PSPC & RP201011$^d$ & 19.7 & 61 & 56 & 4 \\ 
\hline
\end{tabular}
}
\end{center}
{\scriptsize
Notes: Date of observations (a) 2008, (b) 2003, (c) 2004, (d) 1992.
(e) Columns (4) and (5) give respectively the total and the net number 
of detected sources in the field-of-view area, which is given in column (6).
}
\end{table}


\begin{table} 
\caption{Fields chosen for optical spectroscopy.}
\begin{center}
\begin{tabular}{lccccc}
\hline
CMaX & $\alpha_{2000}$ & $\delta_{2000}$ & N$_X$ & N$_c$ & N$_o$\\
\hline
66 & 07:03:53 & -11:34:15 &  16  & 18  & 9 \\ 
71 & 07:03:57 & -11:30:15 &    9  &  10  & 1 \\
75 & 07:04:08 & -11:24:30 &    7  & 12  &  5 \\ 
89 & 07:04:26 & -11:33:45 &    5  &   5  & 23 \\ 
\hline
\label{tab2}
\end{tabular}
\end{center}
{\scriptsize
Notes: N$_X$ is the number of X-ray sources found in each field; N$_c$ is the number of possible optical counterparts, and N$_o$ is the number of other candidates, which are not associated to X-ray sources.
}
\end{table}


\section{Observational Data}

Using the Gemini South telescope, in the multi-object spectroscopy ({\it GMOS}) configuration, we acquired observational data with the good spectral and spatial quality needed to identify the spectral features typical of young stars. A first set of observations using {\it GMOS} in the imaging mode provided the preliminary images to prepare the masks required to the spectroscopy (second set of observations). The {\it GMOS} images were obtained with the {\it R} filter that enables to derive an estimation of optical photometry in this band, as discussed in Sect. 3.2.

\subsection{Gemini multi-object spectroscopy}

Using {\it GMOS} ($\sim$ 5' $\times$ 5' FoVs) pointed to the four stellar groups (Table \ref{tab2}), it was possible to observe a sample of 83
stars that are listed in Table \ref{sample}. Since the {\it XMM} and {\it Chandra} sources around Sh~2-296 are distributed in a 0.25 sq. deg. area that is larger than the {\it GMOS} fields, only 23\% of the total number of X-ray sources were covered in 0.028 sq. deg. area of our observing 
runs (Obs. Id. GS2005B-Q44) that occurred on 2005, November, 8-9.

Figure \ref{fieldtot} (middle and bottom panels) shows the {\it GMOS}  {\it R} band images of the observed FoVs overlapped by contours corresponding to the X-ray emission of the sources detected with {\it XMM-Newton} and the position of the targets for which spectra was acquired. The images are identified by the representative CMaX source found in the respective GMOS field. CMaX~66 is in the same FoV of Z CMa and VdB~92, while the field of CMaX~75 contains BRC~27.


\begin{table*}
\caption{List of observed objects: {\it 2MASS} identification; equivalent width of spectral features (negative values
for emission lines); signal-to-noise ratio; optical {\it R (GMOS)} and {\it I (DENIS)}, and infrared ({\it 2MASS}, {\it WISE}) photometry.}
\begin{center}
\begin{tabular}{ccrccccccccccccc}
\hline
ID & 2MASS ID & WH$\alpha$ (\AA) & WLi (\AA) & S/N & R & I & J & H & K & [3.4] & [4.6]  & [12] & [22]  \\
\hline 
1 &    07034358-1133207     & -13.9 & 0.32 & 11.9 & 13.71 & 16.38 &    14.45:    &    14.10:    &    11.78:    &  &  &  &  \\
2 &    (07034424-1132058)     &  &  & 0.5 & 20.29 &  &  &  &  &  &  &  &  \\
3 &    (07034606-1135120)     & -11.0 &  & 4.1 & 19.90 &  &  &  &  &  &  &  &  \\
4 &    07034651-1135118     & -4.1 &  & 8.1 & 18.65 & 17.19 & 15.70 & 14.94 & 14.70 &  &  &  &  \\
5a  &    07034751-1131489     & -2.2 & 0.55 & 25.5 & 15.63 & 14.88 & 13.52 & 12.77 & 12.60 & 12.11 & 11.75 & 8.86 & 4.59 \\
5b  &    07034782-1131457     &  &  & 14.1 & 15.67 & 16.73 & 15.82 & 15.28 & 14.63 &  &  &  &  \\
6 &    07034869-1131502     & -4.0 & 0.60 & 23.2 & 16.36 & 15.25 & 13.99 & 13.21 & 13.02 &  &  &  &  \\
7 &    07034972-1132169     & -5.8 & 0.56 & 26.7 & 15.20 &  &    14.06:    &    14.29:    &    13.06:    &  &  &  &  \\
8 &    07034994-1132148     & -0.2 & 0.35 & 27.5 & 15.11 & 14.34 &    13.63:    &    13.00:    &    12.73:    &  &  &  &  \\
9 &    07035043-1133425     & -137.2 &  & 2.7 & 19.04 & 17.97 & 15.24 & 14.40 & 13.81 & 13.20 & 12.24 & 9.37 & 6.46 \\
10a  &    07035152-1134557     & 5.3 &  & 9.2 & 11.69 & 11.23 & 10.97 & 10.76 & 10.69 & 10.60 & 10.62 & 7.73 & 5.22 \\
10b  &    07035338-1134504     & -19.0 & 0.45 & 42.7 & 15.52 & 14.33 & 13.44 & 12.48 & 11.93 & 10.71 & 10.35 & 5.97 & 4.03 \\
11 &    07035225-1134195     & -1.9 & 0.51 & 12.6 & 16.74 & 15.70 & 14.51 & 13.62 & 13.50 &  &  &  &  \\
12 &    07035240-1132546     & -1.8 & 0.42 & 22.1 & 17.33 & 16.30 & 15.11 & 14.31 & 14.18 &  &  &  &  \\
13 &    07035303-1129352     & -22.5 & 0.32 & 32.5 & 13.78 & 13.49 & 12.36 & 11.54 & 10.95 & 10.26 & 9.72 & 7.39 & 5.68 \\
14 &    07035400-1132478     & -19.4 & 0.36 & 36.6 & 13.70 & 13.25 & 12.34 & 11.61 & 11.18 & 10.21 & 9.73 & 6.33 & 3.98 \\
15 &    07035414-1128235     & -2.2 & 0.52 & 36.3 & 12.71 &  & 14.12 & 13.43 & 13.31 &  &  &  &  \\
16 &    07035486-1134340     & 4.7 &  & 9.6 & 10.40 & 9.12 & 8.90 & 8.91 & 8.93 & 8.77 & 8.73 & 5.21 & 3.02 \\
17 &    07035503-1128181     & -3.8 & 0.61 & 31.7 & 16.18 & 15.20 & 14.14 & 13.43 & 13.20 &  &  &  &  \\
18 &    07035542-1135149     & -104.6 &  & 9.9 & 15.89 & 14.73 & 12.95 & 11.97 & 11.36 & 10.55 & 10.12 & 6.47 & 4.58 \\
19 &    07035564-1132468     & 2.9 &  & 50.4 & 11.95 & 11.37 & 10.78 & 10.40 & 10.31 & 10.09 & 10.07 & 7.03 & 4.32 \\
20a  &    07035575-1129315     & -1.9 & 0.45 & 35.6 & 14.69 & 13.95 & 12.92 & 12.22 & 12.02 & 11.89 & 11.84 & 11.13 & 8.44 \\
20b  &    07035646-1129476     & -2.8 & 0.44 & 11.2 & 16.55 & 15.51 & 14.03 & 13.24 & 13.01 &  &  &  &  \\
21 &    07035584-1127544     & -2.1 & 0.37 & 35.3 & 15.85 & 15.01 & 14.05 & 13.37 & 13.16 &  &  &  &  \\
22 &    07035587-1133385     &  & 0.38 & 40.5 & 14.83 & 13.82 & 12.37 & 11.54 & 11.24 & 10.90 & 10.88 & 7.58 & 5.51 \\
23 &    07035666-1134553     & -36.3 & 0.37 & 40.7 & 15.58 & 14.77 & 12.94 & 11.87 & 11.17 & 10.01 & 9.39 & 6.02 & 3.87 \\
24 &    07035708-1128317     & -0.5 & 0.46 & 52.2 & 14.10 & 13.44 & 12.37 & 11.69 & 11.47 & 11.35 & 11.34 & 10.16 & 8.40 \\
25 &    07035805-1132398     & -10.6 & 0.54 & 31.4 & 16.32 & 15.49 & 14.10 & 13.23 & 12.93 &  &  &  &  \\
26 &    07035833-1134188     & -6.4 & 0.70 & 8.8 & 17.76 & 16.24 & 14.75 & 13.98 & 13.78 &  &  &  &  \\
27 &    07035880-1135311     & 3.2 &  & 16.5 & 11.42 & 10.97 & 10.72 & 10.54 & 10.47 & 10.25 & 10.23 & 6.80 & 4.42 \\
28 &    07035926-1124234     & 4.3 &  & 13.6 & 16.37 & 16.09 & 15.09 & 14.57 & 14.31 &  &  &  &  \\
29 &    07035994-1130318     & -3.3 & 0.50 & 15.1 & 15.78 & 14.55 &    13.21:    &    12.27:    &    12.25:    & 11.87 & 11.82 & 10.68 & 8.29 \\
30 &    07040041-1133596     & -14.3 & 0.38 & 25.8 & 15.22 & 14.03 & 12.10 & 10.92 & 10.19 & 9.30 & 8.65 & 6.01 & 4.49 \\
31 &    07040114-1136255     & -0.7 & 0.52 & 37.3 & 15.15 & 14.32 &    13.50:    & 12.69 &    12.51:    & 12.31 & 12.24 & 9.28 & 6.41 \\
32 &    07040119-1128454     &  & 0.44 & 15.6 & 15.21 & 14.57 & 13.58 & 12.86 & 12.72 & 12.50 & 12.50 & 10.51 & 8.16 \\
33 &    07040224-1124188     &  em?  &  & 20.2 & 15.29 & 14.86 & 13.47 & 12.64 & 12.46 & 12.20 & 12.14 & 9.71 & 8.37 \\
34a  &    07040225-1125429     & -11.7 &  & 30.2 & 11.40 & 10.38 & 11.31 & 10.75 & 9.94 & 8.28 & 7.49 & 4.10 & 2.27 \\
34b  &    07040234-1125393     & 2.1 &  & 19.2 & 10.49 & 10.38 & 10.40 & 10.32 & 10.26 &  &  &  &  \\
35 &    07040246-1124292     & -6.9 & 0.47 & 37.2 & 14.70 & 14.45 & 13.44 & 12.73 & 12.44 & 11.98 & 11.98 & 10.17 & 8.12 \\
36a  &    07040290-1123375     & -17.4 & 0.47 & 20.9 & 16.11 & 15.49 & 13.56 & 12.43 & 11.86 &  &  &  &  \\
36b  &    (07040302-1123396)     & -10.0 & 0.42 & 13.6 & 16.27 &  &  &  &  &  &  &  &  \\
36c  &    07040314-1123275     & -3.4 &  & 14.7 & 17.04 & 15.92 & 13.03 & 11.57 & 10.69 &  &  &  &  \\
36d  &    07040390-1123480     & -3.1 & 0.56 & 34.9 & 15.43 & 15.15 & 13.93 & 13.10 & 12.84 &  &  &  &  \\
37 &    07040290-1132074     & -27.7 & 0.39 & 30.9 & 15.97 &  & 13.91 & 13.14 & 12.90 &  &  &  &  \\
38 &    07040309-1128071     & -1.6 & 0.64 & 35.5 & 14.92 & 14.26 & 13.30 & 12.63 & 12.49 & 12.35 & 12.37 & 10.77 & 7.64 \\
39 &    07040330-1133586     & -1.9 & 0.53 & 39.7 & 15.75 &  & 13.74 & 12.97 & 12.77 &  &  &  &  \\
40 &    07040393-1126097     & 4.0 &  & 22.0 & 10.36 & 9.74 & 9.76 & 9.72 & 9.63 & 9.52 & 9.56 & 10.34 & 8.29 \\
41 &    07040461-1122328     & -1.0 & 0.47 & 32.9 & 14.59 & 14.46 & 13.43 & 12.69 & 12.51 &  &  &  &  \\
42 &    07040507-1122253     &  &  & 1.4 & 16.60 & 16.26 & 14.67 & 13.91 & 13.62 &  &  &  &  \\
43 &    07040542-1128562     & 2.8 &  & 10.8 & 11.39 & 10.06 & 10.10 & 10.06 & 10.04 & 10.01 & 10.01 & 9.19 & 5.86 \\
44 &    07040725-1123188     & 1.1 & 0.29 & 22.5 & 15.84 & 15.30 & 13.75 & 12.86 & 12.56 &  &  &  &  \\
45 &    07041588-1124055     & 3.1 &  & 11.0 & 10.56 & 9.30 & 8.95 & 8.90 & 8.85 & 8.81 & 8.79 & 8.54 & 7.19 \\
46 &    07041601-1126100     & -7.6 & 0.67 & 10.8 & 14.41 & 14.14 & 12.91 & 12.15 & 11.74 & 11.18 & 10.93 & 9.01 & 7.06 \\
47 &    07041786-1134268     & -6.0 & 0.34 & 12.7 & 16.57 & 15.70 & 14.13 & 13.27 & 13.08 & 12.89 & 12.83 & 11.78 & 8.19 \\
48a  &    07041812-1125280     & -2.6 & 0.64 & 25.6 & 14.30 & 14.17 & 14.39 & 13.56 & 13.39 & 12.13 & 12.11 & 11.56 & 8.07 \\
48b  &    07041839-1125239     & -2.6 & 0.52 & 38.9 & 15.91 & 15.56 & 13.14 & 12.42 & 12.28 & 13.22 & 13.17 & 11.71 & 8.45 \\
49 &    07041840-1133154     & 1.8 &  & 20.0 & 15.90 & 15.59 & 14.72 & 14.21 & 14.14 & 14.26 & 14.43 & 11.92 & 8.59 \\
50 &    07041912-1133480     & -1.4 & 0.51 & 48.7 & 14.72 & 14.27 & 13.10 & 12.42 & 12.16 &  &  &  &  \\
\hline
\label{sample}
\end{tabular}
\end{center}
Notes: Objects with the (a), (b), (c) and (d) notation are associated with the same X-ray emitting source. Coordinates indicated between parenthesis are used to identify the three objects without {\it 2MASS} counterpart. Among the {\it 2MASS} sources there are 6 with bad quality data, which are indicated
by  ``:" appearing together with the given value.
\end{table*}


\begin{table*}
\contcaption{List of observed objects: {\it 2MASS} identification; equivalent width of spectral features (negative values
for emission lines); signal-to-noise ratio; optical {\it R (GMOS)} and {\it I (DENIS)}, and infrared ({\it 2MASS}, {\it WISE}) photometry.}
\begin{center}
\begin{tabular}{ccrccccccccccccc}
\hline
ID & 2MASS ID & WH$\alpha$ (\AA) & WLi (\AA) & S/N & R & I & J & H & K & [3.4] & [4.6]  & [12] & [22]  \\
\hline
51 &    07042012-1134040     & 0.7 &  & 14.9 & 12.87 & 12.58 & 11.88 & 11.51 & 11.41 & 11.24 & 11.24 & 11.03 & 8.08 \\
52 &    07042074-1133314     & 3.0 &  & 29.7 & 15.13 & 14.80 & 14.02 & 13.60 & 13.47 & 13.29 & 13.41 & 11.84 & 8.53 \\
53 &    07042217-1135036     & 2.4 &  & 41.9 & 15.36 & 15.05 & 14.44 & 14.10 & 13.94 & 13.75 & 13.61 & 10.85 & 7.27 \\
54 &    07042316-1134320     & 4.4 &  & 0.9 & 15.10 & 14.54 & 13.22 & 12.47 & 12.23 & 12.09 & 12.11 & 10.79 & 7.87 \\
55 &    07042340-1132515     & -1.7 & 0.50 & 30.4 & 14.96 & 14.32 & 13.22 & 12.48 & 12.32 & 12.14 & 12.08 & 10.10 & 6.80 \\
56 &    07042383-1136135     & 5.0 &  & 11.2 & 13.77 &  & 13.03 & 12.76 & 12.70 & 12.61 & 12.64 & 12.12 & 8.76 \\
57 &    07042427-1135183     & 2.4 &  & 11.0 & 11.66 & 11.34 & 10.43 & 9.85 & 9.68 & 9.58 & 9.62 & 9.56 & 8.26 \\
58 &    07042429-1132190     & 2.1 &  & 47.1 & 17.16 & 16.78 & 15.92 & 15.74 & 14.99 & 16.12 & 15.98 & 11.41 & 8.48 \\
59 &    07042581-1133235     &  &  & 0.0 & 13.49 & 13.42 & 13.09 & 12.92 & 12.91 & 12.76 & 12.80 & 11.98 & 8.41 \\
60 &    07042625-1131207     & -46.2 &  & 0.7 & 13.38 & 13.72 & 12.47 & 11.60 & 11.15 & 10.67 & 10.28 & 8.19 & 5.87 \\
61 &    07042640-1132211     & 6.7 &  & 23.4 & 11.92 & 11.95 & 11.70 & 11.60 & 11.59 & 11.49 & 11.51 & 12.40 & 8.58 \\
62 &    07042736-1135241     & 3.7 &  & 22.5 & 16.01 & 15.70 & 15.04 & 14.73 & 14.58 &  &  &  &  \\
63 &    07042798-1134431     & -0.2 & 0.20 & 33.1 & 13.83 & 13.43 & 12.41 & 11.73 & 11.53 & 11.40 & 11.37 & 11.98 & 8.24 \\
64 &    07042834-1133379     & 0.9 &  & 23.4 & 14.43 & 14.11 & 13.35 & 12.79 & 12.65 &  &  &  &  \\
65 &    07042910-1135294     & 1.3 &  & 51.5 & 15.26 & 14.72 & 13.48 & 12.73 & 12.58 & 12.36 & 12.37 & 11.69 & 8.65 \\
66 &    07042965-1135592     & 1.6 &  & 65.4 & 13.27 & 13.58 & 12.71 & 12.37 & 12.30 & 12.17 & 12.23 & 11.36 & 8.31 \\
67 &    07042978-1136041     & 2.7 &  & 29.3 & 13.57 & 13.22 & 16.42 &    14.30:    & 13.85 &  &  &  &  \\
68 &    07043025-1134211     & 1.8 &  & 31.1 & 14.52 & 14.25 & 13.56 & 13.14 & 13.00 & 12.97 & 13.02 & 12.54 & 8.89 \\
69 &    07043056-1132397     & 2.9 &  & 53.4 & 15.41 & 14.98 & 14.25 & 13.80 & 13.76 &  &  &  &  \\
70 &    07043319-1131430     & 2.5 &  & 68.6 & 16.13 & 15.86 & 15.05 & 14.76 & 14.73 &  &  &  &  \\
71 &    07043406-1133054     & 4.2 &  & 22.0 & 12.80 & 12.64 & 12.15 & 12.00 & 11.82 & 11.76 & 11.81 & 12.62 & 8.42 \\
72 &    07043428-1134501     & -3.1 & 0.34 & 13.8 & 16.64 & 16.00 & 14.43 & 13.66 & 13.45 &  &  &  &  \\
73 &    07043534-1134121     & 3.8 &  & 11.5 & 13.11 & 12.98 & 12.35 & 12.02 & 11.96 & 11.78 & 11.76 & 12.37 & 8.43 \\
74 &    07043596-1133082     & 1.4 &  & 58.0 & 12.89 & 12.53 & 11.49 & 10.88 & 10.67 & 10.54 & 10.61 & 11.01 & 7.70 \\
75 &    07043666-1131496     & 3.0 &  & 42.8 & 14.70 & 14.48 & 13.83 & 13.54 & 13.37 & 13.24 & 13.12 & 10.33 & 5.17 \\
\hline
\label{sample2}
\end{tabular}
\end{center}
\end{table*}

The high sensitivity of the Gemini telescope allows us to resolve possible binary systems and to identify the best candidate to be the optical counterpart of the X-ray source in these cases. It was also possible to investigate the presence or not of a faint star associated with a given source that does not have any optical counterpart identified in the published catalogues.

For spectroscopy we used the R600/652 and R600/648 grating configurations that provide the adequate spectral resolution  (R$\sim$3500) and cover the spectral range ($\sim$ 5000 - 8000 $\AA$) required to detect typical features of young stars (H$\alpha$, Lithium). The choice of two central wavelengths provides an overlap between the two sets of spectra obtained for each field, allowing the detection of features that would be missing due to the CCD gaps. Long and short time exposures (10 and 3 minutes) were applied in order to avoid sub-exposition of faint objects and/or saturation of bright stars.

\subsection{Optical and infrared photometry}

With {\it GMOS} in imaging mode we obtained {\it R} (630nm)  magnitude of the targets, which are typically 13 - 18 mag. A rough photometry was estimated
during the mask preparation that uses the {\it IRAF-GMOS} standard  procedures.

To complement the photometry, we searched in the literature the data covering bands from the optical to the mid-IR.

The catalogues {\it NOMAD} \citep{nomad}, {\it GSC} \citep{gsc} and {\it DENIS}\footnote{DENIS Consortium, 2005 - http://cds.u-strasbg.fr/denis.html} were inspected to retrieve the {\it BVRI} photometry.  A calibration of the {\it GMOS} photometry was obtained by comparing our data with the {\it R} magnitude from {\it NOMAD} database.  Three of the fields showed a good correlation between $R_{GMOS}$ and $R_{NOMAD}$ within
a 2$\sigma$ dispersion, where $\sigma$=0.2 mag. A systematic deviation of 2.5 mag was found only for field CMaX75. This value was used to correct the $R_{GMOS}$  of all the stars of this field leading to the same correlation found for the other fields.

Near-infrared data at the {\it J}, {\it H} and {\it K} bands were extracted  from the Two Micron All-Sky Survey - {\it 2MASS} - \citep{2mass} and the mid-Infrared data at 3.4, 4.6, 12, and 22 $\mu$m are from the Wide-field Infrared Survey Explorer - {\it WISE} \citep{wise}. 

Table \ref{sample} gives the photometric data that were used in the present work. Considering the lack of good quality data in the {\it B} and {\it V} band for most of the objects in our sample, optical photometry is represented by the {\it R} ({\it GMOS}) and {\it I} ({\it DENIS}) bands only.


\section{Spectroscopic classification}

\begin{figure}
\begin{center}
\includegraphics[width=8cm,angle=0]{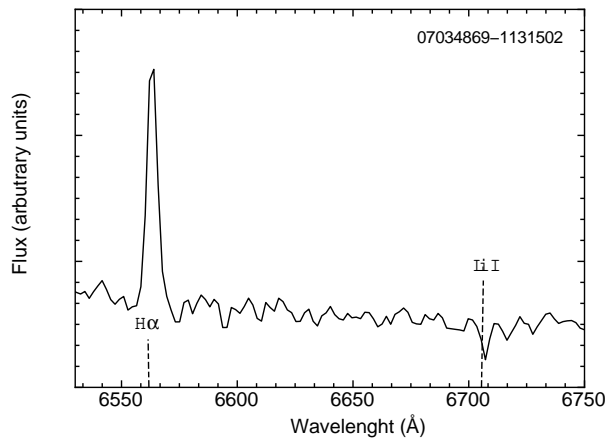}
\caption{ Example of a cut of one of the observed spectra covering the region between 6500 and 6750 \AA, where the H$\alpha$ emission and the lithium absorption lines can be seen. }
\label{lines}
\end{center}
\end{figure}

To identify the young stars in our sample we inspected the {\it GMOS} spectra, searching for typical features such as H$\alpha$ and Li I lines, which are illustrated in Fig. \ref{lines}. The equivalent widths 
W(H$\alpha$) and W(Li) are listed in Table \ref{sample}.

 The occurrence of H$\alpha$ emission and/or Li I lines in the spectra observed for our sample was used as criterion to reveal the presence of 41 T Tauri stars. Among the other 42 objects, which do not fulfil these spectral criteria, there are 17 young star candidates that show one or more of the following criteria: H$\alpha$ emission line; counterpart of X-ray source; association with Sh~2-296 reported in the literature (Herbst, Racine and Warner 1978, Schevchenko et al. 1999). The additional 25 observed stars, which have no typical spectral feature, depend on the IR characterization to have their nature investigated. In summary, our sample of 83 observed stars contains three groups: 41 confirmed T Tauri stars, 17 young star candidates, and 25 additional objects that were included in the IR analysis (Sect. 5.1). Details about these additional objects are discussed in Appendix A.

Our classification is based on different approaches: (i) the estimation of spectral types by comparing the observed the continuum shape and TiO band with library templates; (ii) the separation of CTTs from WTTs, based on W(H$\alpha$); (iii) the comparison of our sample with other young stellar groups providing a qualitative determination of age, based on the lithium depletion. The results of the spectroscopic classification are described in the next sub-sections.

\subsection{Spectral types}

The comparison of the observed continuum with spectra libraries is useful to roughly determine
the spectral type of the optical counterparts. In particular,  the TiO bands (7050-7150 \AA), which are prominent features in K and M-type stars  \citep{Mortier2011}, were
used to refine the spectral classification of the sample.  

\begin{figure}
\begin{center}
\includegraphics[width=6.5cm,angle=0]{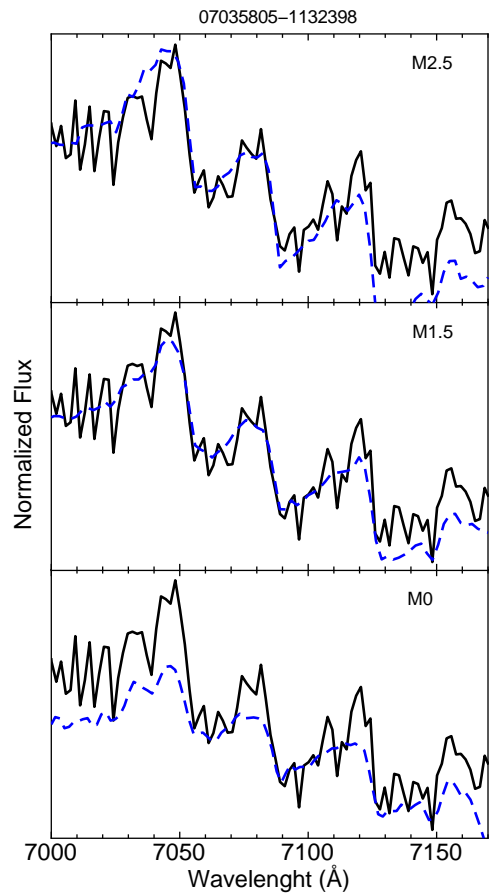}
\caption{Examples of the spectral classification using the TiO bands between 7050 and 7150 \AA \  for object 07035805-1132398 (black line), which was classified as a M1.5 type star. The library spectra for spectral types M0, M1.5 and M2.5 are showed by dashed blue lines, for comparison.
 }
\label{sptype} 
\end{center}
\end{figure}

We performed the spectral typing by determining the best match of the continuum shape, fitting our science spectra to a grid of templates. For this fitting, we adopted  the STELIB spectral library  \citep{LeBorgne2003} and the {\it M Dwarf and Giant Spectral Standards} made available by Kelle Cruz\footnote{http://kellecruz.com/M\_standards}.  The grid of compiled library spectra goes down to spectral type M9 in steps of one spectral type, or 0.5 spectral types in few cases. The TiO bands were used to refine the spectral classification, giving a typical uncertainty of one spectral type, as illustrated in Fig. \ref{sptype}, where we show a star from our sample classified as M1.5, superimposed with the library spectra of same spectral type and varying in the interval of 1 spectral type for comparison.

To estimate the effective temperatures we used the spectral type $vs.$ temperature scale from \cite{Luhman2003}, for spectral types from M0 to M6, and the models from \cite{Siess2000}, for spectral types from K0 to K7. The spectral types and effective temperatures determined in the present work are listed in Table \ref{ysos}.

\cite{Bayo2011} compared the \cite{Luhman2003}  temperature scale to their own and those from other works. We note differences of  $\sim$ 250 K between the scales, which would cause errors larger than those derived from our spectral type determinations. Therefore we assume 250~K as the average uncertainty for the effective temperatures.

\subsection{H$\alpha$ emission}

The H$\alpha$ line can be used to  distinguish CTTs from WTTs. These two classes of objects are similar in age but have circumstellar structures showing different evolutionary characteristics \citep{White2003}. While in WTTs the H$\alpha$ emission is due only to the star chromospheric activity, in CTTs broder lines are believed to be a result of the star undergoing active accretion from a circumstellar disc.

\cite{White2003} found an empirical criterion to distinguish the T Tauri classes. Assessing the value of W(H$\alpha$) they noted that objects presenting veiling, i.e. CTTs, are stars of spectral type: (i) K0 - K5 with W(H$\alpha$) $\geq$ 3\AA, (ii) K7 - M2.5 with W(H$\alpha$) $\geq$ 10 \AA, (iii) M3 - M5.5 with W(H$\alpha$) $\geq$ 20 \AA \ and (iv) M6 - M7.5 with W(H$\alpha $) $\geq$ 40 \AA.

\begin{figure}
\centering
\includegraphics[width=8.5cm,angle=0]{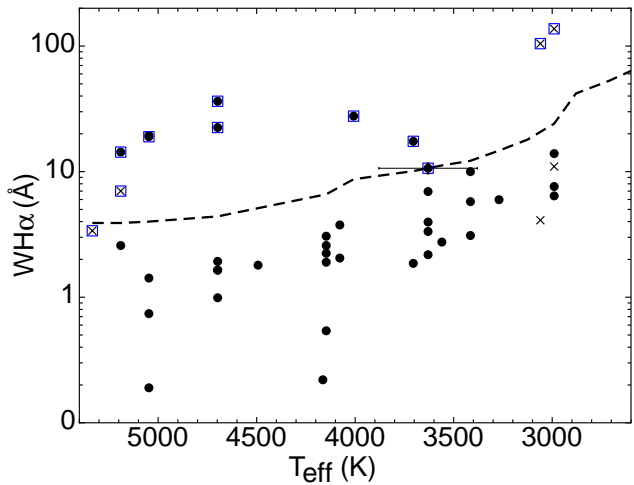}
\caption{Equivalent width of the H$\alpha$ line as a function of spectral type of our sample of T Tauri stars (filled circles) and young star candidates (crosses)  showing H$\alpha$ emission. The dashed line represents the activity limit found in young star clusters, objects above that threshold are suggested to be CTTs according to Barrado y Navascu\'es \& Mart\'in (2003). Stars also classified as CTTs according to the empirical criterion proposed by White \& Basri (2003) are highlighted by blue squares.}
\label{criterios}
\end{figure}

A similar approach is proposed by \cite{Barrado2003}, considering the maximum H$\alpha$ flux for objects found in young stellar clusters. In Fig. \ref{criterios} we present the results of applying both of these criteria to our sample, by comparing W(H$\alpha$) with T$_{eff}$ for the objects showing   H$\alpha$ emission  (W(H$\alpha$) $<$ 0 in Table 3)  and having well determined spectral types.  It is noted in Fig. \ref{criterios} that the two criteria are equivalent, considering the uncertainties. 

We were able to estimate the spectral type for the 41 T Tauri stars identified with basis on H$\alpha$ and Li (see Sec. 4.2), which are classified as: 7 CTTs and 34 WTTs. Among the other emission line stars that were included in the list of young star candidates, there are 6 for which we estimated the effective temperature (from spectral type). These candidates were also evaluated according the criterion used in Fig. \ref{criterios}, indicating that four of them probably are CTTs and two could be WTTs, but the confirmation of their nature depends on complementary analysis. 

\subsection{Youth indicated by the lithium depletion}

The presence of lithium in the stellar spectrum is a valuable indicator of youth, since it is rapidly destroyed as the star approaches the main-sequence.

\begin{figure}
\centering
\includegraphics[width=8.5cm,angle=0]{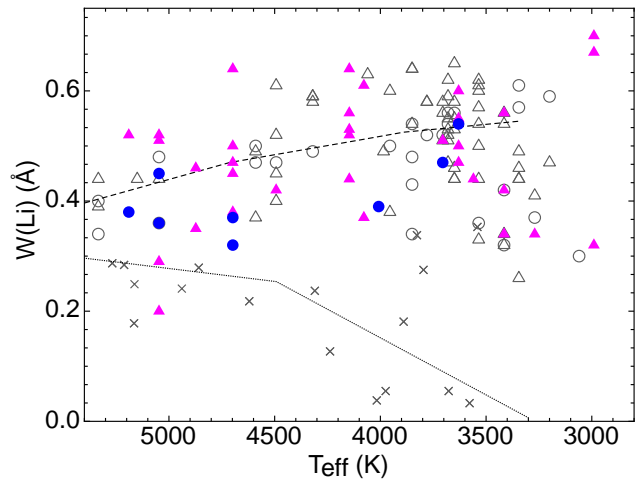}
\caption{ Equivalent width of Li I line vs. effective temperature. Blue filled circles and magenta filled triangles represent CTTs and WTTs in our sample, respectively, while empty symbols show CTTs and WTTs members of the L1641 cluster for which Fang et al. (2013)  estimated an average age of 1.5 Myr (dashed line).  Crosses indicate the objects of the Tucanae-Horologium association, which has an average age of 27 Myr (Mentuch et al. 2008) represented by the dotted line. }
\label{litio}
\end{figure}

Lithium abundance is clearly observed to systematically decrease with age when groups of stars in different evolutionary stages are compared (Mentuch et al. 2008, da Silva et al. 2009, Fang et al. 2013). In Fig. \ref{litio} we show W(Li) as a function of spectral type for WTTs and CTTs in our sample, we also show the mean value of W(Li) for young stars in L~1641 corresponding to an age of 1.5 Myr \citep{Fang2013} and objects in the Tucanae-Horologium association with an age of 27 $\pm$ 11 Myr \citep{Mentuch2008}. 

The distribution of W(Li) in our sample is similar to the one showed by the population in L~1641, making it consistent with an age of at least 1 Myr. However a large scatter in values  of W(Li) is observed. This could be related to an age spread, but is more likely connected to other factors such as different rotation rates or accretion history \citep{daSilva2009, Baraffe2010}.
 
As noted for the Tucanae-Horologium sample in Fig. \ref{litio}, lithium depletion appears to be sensitive to mass in older groups, since there is a decline on lithium abundance for late spectral types (Mentuch et al. 2008, da Silva et al. 2009, Murphy et al. 2013), which is not observed in our sample nor in L~1641.

\begin{figure*}
\centering
\includegraphics[width=16cm,angle=0]{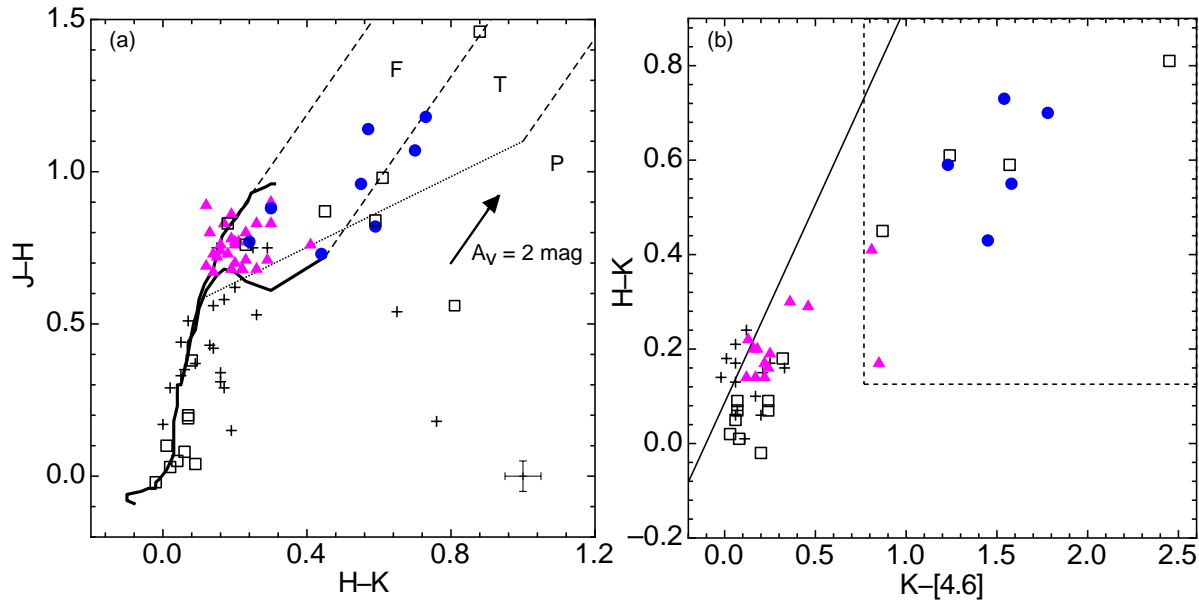}
\caption{{\it (a)}: 2MASS colour-colour diagram for our sample. CTTs and WTTs are respectively represented by filled circles and triangles. Open squares show the PMS star candidates and crosses indicate the  unclassified additional stars. The locus for the main-sequence and giants (solid curves) are from Siess et al. (2000) and Bessell \& Brett (1988), respectively. T Tauri stars are expected to be found above the dotted line (Meyer et al. 1997). Dashed lines represent the reddening vectors. {\it(b)}: 2MASS - WISE diagram. The thick line corresponds to the redening vectors, using the extinction law by Flaherty et al. (2007). The dashed line delimits the region occupied by PMS stars in the sample of Cusano et al. (2011). 
}
\label{diagrams}
\end{figure*}

\section{Photometric Characterization}

In addition to the optical spectroscopic criteria adopted to classify the young stars, we performed an analysis using the near- and mid-IR data available for our sample.  The IR characterization is based on colour-colour diagram aiming to evaluate reddening and IR excess (Sect. 5.1). We also searched for indications of the presence of circumstellar disc, by using a genetic algorithm  code to fit the observed spectral energy distribution (Sect. 5.2).

\subsection{Infrared colours}

Figure \ref{diagrams}a shows the {\it J-H  vs. H-K} diagram for 74 objects of our sample having good 
quality {\it 2MASS} data,  52 of them are confirmed young stars and candidates, while the other 22 
are additional objects (Appendix A). The solid curves represent the main-sequence
and giant stars \citep{Bessell1988}. The dashed lines represent the reddening vectors and the dotted line indicates the expected locus of T Tauri stars \citep{Meyer1997}. Following \cite{Jose2011} we use three regions (F, T and P) in the diagram to classify the objects. The F region is bounded by the reddening vectors for main-sequence stars, here we find  reddened field stars and young stars with little or no excess in the near-IR (usually WTTs,  but CTTs  may also be present). The objects found to the right of the F region show near-IR excess. In the T region, in particular, we expect to find mostly CTTs with high near-IR excess. In the P region we can find Class I objects.

It can be noted in Fig. \ref{diagrams}a that most of the additional objects follow the trend of the main-sequence curve, while the WTTs identified in our sample fall in the the region expected for young objects with low near-IR excess. The objects identified as CTTs tend to show at least some near-IR excess, which is consistent with their classification since it is indicative of the presence of circumstellar material.

In the  {\it H-K  vs. K-[4.6]} diagram of Fig. \ref{diagrams}b, we combine data from the {\it WISE} and {\it 2MASS} catalogues. As in the {\it 2MASS}  colour-colour diagram, the location of objects on the right side of the reddening vector is an indication of excess in the mid-IR. We adopted the same extinction law given by 
\cite{Flaherty2007} for  IRAC 2 (4.5 $\mu$m) band. As the difference between this band and WISE W2 (4.6 $\mu$m) is minimal, it can be ignored \citep{Scholz2013}.

We also highlight in Fig. \ref{diagrams}b the region where \cite{Cusano2011} found a concentration of reddened PMS objects identified in their study of star forming region Sh~2-284. Compared to our sample, we see that  this region coincides with the location of the CTTs classified by us. On the other hand, most of the stars classified as WTTs fall along the reddening vector.

The extinction affecting our sample was determined using the observed and intrinsic {\it J-H} colours, derived from the spectral types, to estimate $A_K=([J-H]_{obs}-[J-H]_o)/0.95$  \citep{Fang2013}. We adopted a normal total-to-selective extinction value (R$_V$=3.1) and the extinction law of \cite{Cardelli1989}.
Table 4 gives the $A_V$ obtained for the objects having both, spectral type and near-IR data. 
On the lack of these data, we assumed a mean value of 1 mag for the visual extinction.

\begin{figure*}
\centering
\includegraphics[width=18cm,angle=0]{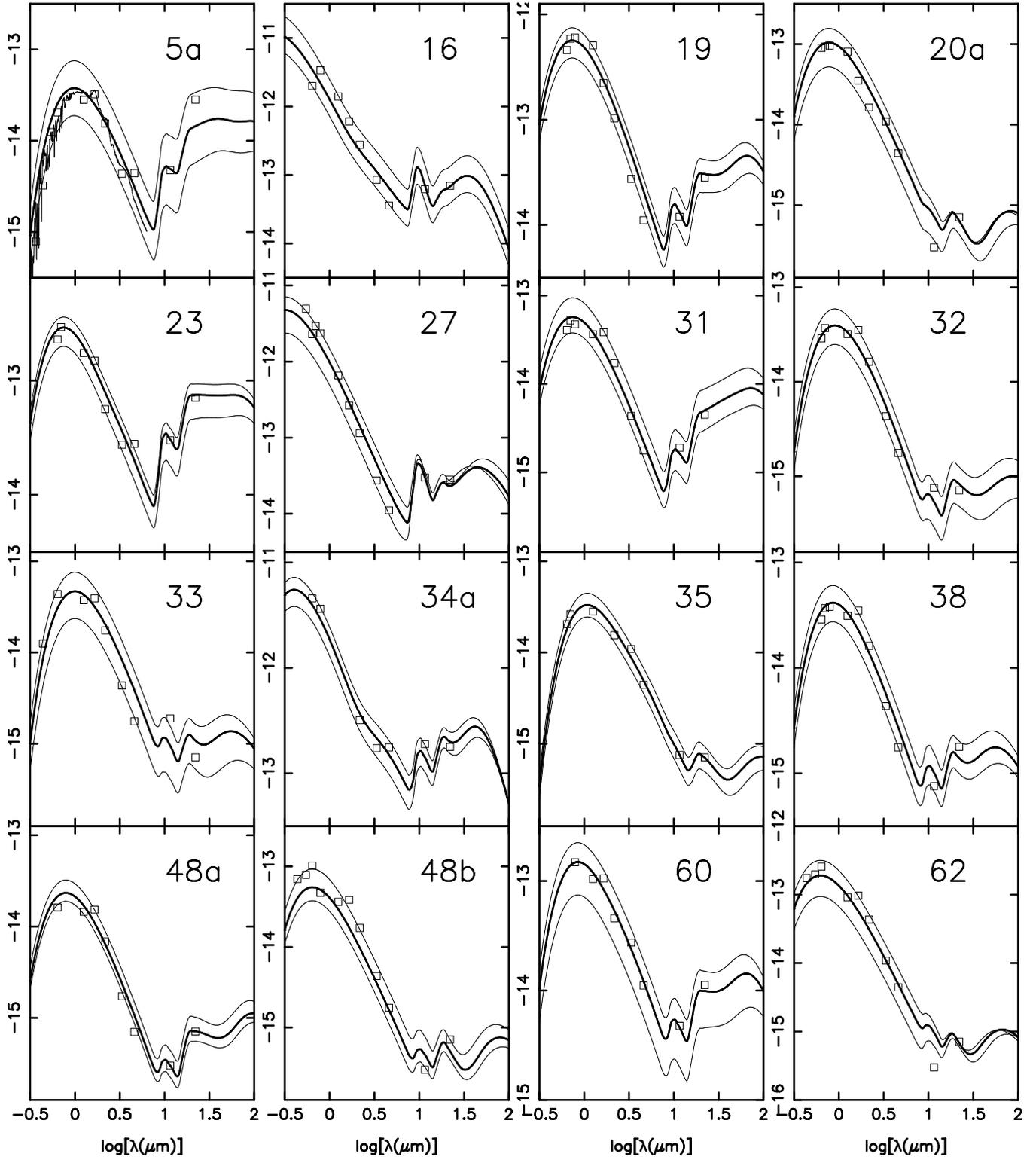}
\caption{ Spectral energy distribution of PMS stars. The open squares represent the dereddened flux at each band in units of Watt m$^{-2}$.  The thick line represents the best SED fitting, while thin lines show minimum and maximum variation. An example of the Kurucz model for normal photospheres of the same spectral type as the star is shown for object No. 5a.}
\label{seds}
\end{figure*}

\subsection{SED fitting}

One of the explanations for the infrared excess is the dust emission due to the presence of a circumstellar disc.  
Aiming to estimate a possible circumstellar emission in our sample, we used a disc model to fit the observed \cite{Hetem2007} that uses genetic algorithm to optimize the parameter estimation. GADisk is based on 
the flared disc configuration suggested by \cite{Dullemond2001} for a passively irradiated circumstellar disc 
with an inner hole that is formed by photo-evaporation due to the high temperatures close to the central star. Different disc components contribute to the total emitted flux in addition to the stellar 
emission. The circumstellar  contribution can be expressed as a fraction of the integrated observed flux, defined by  $f_c$ = ( F$_{total}$ - F$_{star}$ ) / F$_{total}$.  

Figure \ref{seds} shows examples of the synthetic reproduction of the observed SEDs. 
Three curves calculated by GADisk illustrate the expected variation in the estimation of integrated flux, 
where the thick line indicates the best fitting. In the case of source No. 5a, for example, $f_c$ is 46\% with lower and upper levels varying from 43\% to 47\%, 
which is the typical dispersion found in our sample, leading us to adopt an error of 6\% in the estimation of $f_c$.

The parameters obtained from the SED fitting: stellar temperature (T$_*$), visual extinction (A$_{Vsed}$), and $f_c$, are given in Table 4. It is interesting 
to note that these results can also be used as an independent confirmation of the spectral classification and the 
adopted extinction.
By assuming a deviation of $\sigma$ = 250 K on the effective temperature derived from spectral type, we verified that among the 23 SED fittings, 18 stars show T$_*$ in agreement with T$_{eff}$ within 1$\sigma$ deviation. Only 5 objects (No. 14, 30, 38, 48a,b) have differences of about 2$\sigma$ on the temperatures that are probably due to uncertainties of the spectral type derivation for these objects. The same can be said about the differences among the visual extinction obtained from the spectral analysis (Sect. 5.1) and SED fitting, which are less than 0.5 mag for 87\% of the stars. Larger differences are found for objects No. 30, 31 and 47.

For comparison, Fig. \ref{seds} also shows an example of the photosphere emission calculated with the Kurucz model by adopting the stellar temperature that gives the best fitting of the un-reddened observed data, which  coincides with the effective temperature derived from spectral type.

Significant circumstellar emission ($f_c > 30$\%) is found for 7 objects: 1 HAe, 1 WTT, and 5 CTT stars, which also show IR excess (K-[4.6]$>$ 0.8) in the colour-colour diagram of Fig. \ref{diagrams}b. Even if the SED fitting could not be performed for the whole sample (due to the lack of mid-IR photometry), we can guess that those objects not detected by WISE probably do not show significant IR excess neither circumstellar emission. In this case, we suggest that among  the 41 confirmed TT stars 75\% can be considered discless.

A discussion on the age distribution of discless and disc-bearing stars is presented in Sect. 6.1.


\begin{table*}
\caption{ List of young stars and candidates. The X-ray emission is relative to the minimum value of 0.7 counts/ks detected with {\it XMM-Newton} for this sample. Spectral types types and visual extinction were derived from the spectroscopic analysis. Stellar temperature (T$_*$), A$_{Vsed}$ and f$_c$ are results from the SED fitting. Approximate ranges of ages and masses are derived from Fig \ref{cmd}. The last column lists other references for the objects found in the literature.}
 \begin{tabular}{ccccccccccccc}
  \hline
ID & F$_X$ & Sp. T. & T$_{eff}$  & A$_V$ & T$_*$ & A$_{V_{SED}}$ & f$_c$ & IR-e & age & mass & Class. & Refs. \\
     &            &           & (K)            & (mag)   & (K)      & (mag)                & (\%)   &  & (Myr) & (M${\odot}$) & & \\
\hline
1 &  &  M6  & 2990 &  &  &  &  &  &  &  &   WTT   &  \\
3 &  &  M6  & 2990 &  &  &  &  &  &  &  &  Me  &  \\
4 &  &  M5.5  & 3060 & 1.1 &  &  &  &  &$>$ 15 & 0.2 - 1 &  Me &  \\
5a & 4.0 &  M1.5  & 3630 & 0.7 & 3750 & 0.5 & 46 &  & 1 - 2 & 0.2 - 1 &   WTT   &  \\
6 & 2.6 &  M1.5  & 3630 & 0.9 &  &  &  &  & 1 - 2 & 0.2 - 1 &   WTT   &  \\
7 & 2.4 &  M3  & 3415 &  &  &  &  &  &  &  &   WTT   &  \\
8 & 3.0 &  K3  & 4873 & 0.9 &  &  &  &  &  &  &   WTT   &  \\
9 &  &  M6  & 2990 & 1.5 & 3000 & 1.5 & 29 & T,W & 8 - 10 & 0.2 - 1 &  CTT?  &  \\
10a & 14.6 &  &  &  & 11259 & 2.0 & 1 &  &  &  &  PMS  &   [HRW] 11a / CMaX66       \\
10b &  &  K2  & 5047 & 3.9 & 4900 & 2.0 & 38 & F, W & 2 - 5 & 1 - 2 &   CTT   &  \\
11 &  &  M1  & 3705 & 1.8 &  &  &  &  & 1 - 2 & 0.2 - 1 &   WTT   &  \\
12 &  &  K5  & 4493 & 1.7 &  &  &  &  &$>$ 15  & 0.2 - 1 &   WTT   &  \\
13 & 3.6 &  K4  & 4698 & 2.2 & 4867 & 2.5 & 11 & T, W & 1 - 2 & 1 - 2 &   CTT   &  \\
14 & 3.1 &  K2  & 5047 & 2.0 & 6102 & 2.0 & 33 & F, W & $\sim$ 2 & 1 - 2 &   CTT   &  \\
15 & 1.0 &  K7  & 4147 & 0.4 &  &  &  &  & 5 - 8 & 0.2 - 1 &   WTT   &  \\
16 & 8.4 &  &  &  & 15212 & 1.0 & 2 &  &  &  &  PMS  &    [SEI99] 91        \\
17 & 2.3 &  K7/M0  & 4078 & 0.5 &  &  &  &  & 2 - 5 & 0.2 - 1 &   WTT   &  \\
18 & 1.0 &  M5.5  & 3060 & 2.9 & 3285 & 3.0 & 33 & T, W & 0.1 - 1 & 0.2 - 1 &   CTT?  &  \\
19 & 3.1 &  &  &  & 8039 & 2.0 & 3 &  & 2 - 5 & 3 - 4 &  PMS  &  \\
20a & 7.3 &  K4  & 4698 & 1.3 & 4750 & 1.0 & 2 &  & 2 - 5 & 1 - 2 &   WTT   &   CMaX71      \\
20b &  &  M2  & 3560 & 1.0 &  &  &  &  & 1 - 2 & 0.2 - 1 &   WTT   &  \\
21 &  &  K7/M0  & 4078 & 0.2 &  &  &  &  & 1 - 2 & 0.2 - 1 &   WTT   &  \\
22 & 2.2 &  K4  & 4698 & 2.4 & 4574 & 3.0 & 14 &  & 1 - 2 & 1 - 2 &   WTT   &  \\
23 &  &  K4  & 4698 & 4.3 & 4900 & 4.0 & 33 & T, W & 1 - 2 & 1 - 2 &   CTT   &  \\
24 & 4.0 &  K3  & 4873 & 1.3 & 4821 & 1.0 & 1 &  & 2 - 5 & 1 - 2 &   WTT   &  \\
25 & 1.0 &  M1.5  & 3630 & 1.7 &  &  &  &  & 1 - 2 & 0.2 - 1 &   WTT   &  \\
26 & 1.3 &  M6  & 2990 & 0.8 &  &  &  &  & 5 & 0.2 - 1 &   WTT   &  \\
27 & 4.6 &  &  &  & 11500 & 2.0 & 2 &  & 1 - 2 & $>$ 4 &  PMS  &   [SEI99] 95        \\
29 & 2.3 &  M1.5  & 3630 & 2.2 & 3800 & 1.8 & 3 &  &  &  &   WTT   &  \\
30 & 3.0 &  K1  & 5190 & 6.0 & 4590 & 4.0 & 35 & T, W & $\sim$1 & ~3 &   CTT   &  \\
31 &  &  K2  & 5047 & 2.6 & 4935 & 1.5 & 18 &  &  &  &   WTT   &  \\
32 & 1.9 &  K7  & 4147 & 0.7 & 4076 & 1.0 & 8 &  & 2 - 5 & 0.2 - 1 &   WTT   &  \\
33 &  &  M3  & 3415 & 1.6 & 3693 & 1.6 & 6 &  & 1 - 2 & 0.2 - 1 &  TT?  &  \\
34a & 11.6 &  &  &  & 10436 & 4.0 & 36 & P, W & 1 - 2 & $>$ 4 &   H Ae?   &   [OSP2002] BRC27 33         \\
34b &  &  &  &  & 15010 & 1.0 & 0 &  & 1 - 2 & $>$ 4 &  PMS  &   [SEI99] 99 / CMaX75        \\
35 &  &  M1.5  & 3630 & 0.3 & 3477 & 0.5 & 8 &  & 1 - 2 & 0.2 - 1 &   WTT   &  \\
36a & 22.3 &  M1  & 3705 & 3.8 &  &  &  & F, W & 1 - 2 & 0.2 - 1 &   CTT   &  [CPO2009] 84       \\
36b &  &  M3  & 3415 &  &  &  &  &  &  &  &   WTT   &  \\
36c &  &  K0  & 5334 & 8.5 &  &  &  & T, W & $\sim$1 & ~ 3 &  CTT?  &   [CPO2009]  107        \\
36d &  &  K7  & 4147 & 1.6 &  &  &  &  & 2 - 5 & 0.2 - 1 &   WTT   &  \\
37 & 3.1 &  M0  & 4008 & 0.9 &  &  &  &  & 2 - 5 & 0.2 - 1 &   CTT   &  \\
38 & 4.3 &  K4  & 4698 & 1.0 & 4278 & 1.7 & 7 &  & 5 - 8 & 1 - 2 &   WTT   &  \\
39 & 1.3 &  K7  & 4147 & 1.1 &  &  &  &  & 2 - 5 & 0.2 - 1 &   WTT   &  \\
40 & 8.1 &  &  &  & 8640 & 0.0 & 0 &  & 1 - 2 & $>$ 4 &  PMS  &   [SEI99] 102         \\
41 &  &  K4  & 4698 & 1.5 &  &  &  &  & 5 - 8 & 1 - 2 &   WTT   &  \\
43 & 2.7 &  &  &  & 10000 & 0.0 & 0 &  & 1 - 2 & $>$ 4 &  PMS  &   [SEI99] 103         \\
44 & 11.0 &  K2  & 5047 & 3.3 &  &  &  &  & 5 - 8 & 1 - 2 &   WTT   &  \\
45 & 8.3 &  &  &  & 13000 &  1.5 & 1 &  & 1 - 2 & $>$ 4 &  PMS  &   [SEI99] 114 / CMaX82         \\
46 & 11.4 &  M6  & 2990 & 0.8 & 2800 & 0.8 & 8 & F, W & 1 - 2 & 0.2 - 1 &   WTT   &  \\
47 &  &  M4  & 3270 & 2.0 & 3532 & 0.0 & 15 &  & 2 - 5 & ~ 1 &   WTT   &  \\
48a & 10.9 &  K7  & 4147 & 1.6 & 4600 & 1.5 & 9 &  & 2 - 5 & 0.2 - 1 &   WTT   &  \\
48b &  &  K1  & 5190 & 2.1 & 5718 & 1.5 & 3 &  & 5 - 8 & 1 - 2 &   WTT   &  \\
50 & 3.1 &  K2  & 5047 & 1.5 &  &  &  &  & 5 - 8 & 1 - 2 &   WTT   &  CMaX86       \\
55 & 16.0 &  K4  & 4698 & 1.6 & 4532 & 1.0 & 14 &  & 2 - 5 & 1 - 2 &   WTT   &  \\
60 & 9.3 &  &  &  & 4330 & 2.0 & 15 & F, W &  &  &  CTT?  &  CMaX89       \\
61 &  &  &  &  & 10245 & 1.0 & 0 &  & 1 - 2 & $>$ 4 &  PMS  &   [SEI99] 121        \\
63 & 25.7 &  K2  & 5047 & 1.5 & 5225 & 1.5 & 2 &  & 2 - 5 & 1 - 2 &   WTT   &  \\
72 & 1.4 &  M3  & 3415 & 1.1 &  &  &  &  & 5 - 8 & 0.2 - 1 &   WTT   &  \\

\hline
\label{ysos}
\end{tabular}

\end{table*}


\section{Analysis}

 Based on the IR photometry, we evaluated the ranges of masses and ages, traced by the near-IR excess, which are described in Sect. 6.1.
The photometric characteristics and SED fitting are also used to classify the candidates of our sample (Sect. 6.2), and the spatial 
distribution of different classes is discussed in Sect. 6.3.

\subsection{Age and Mass}

\begin{figure}
\centering
\includegraphics[width=0.47\textwidth,angle=0]{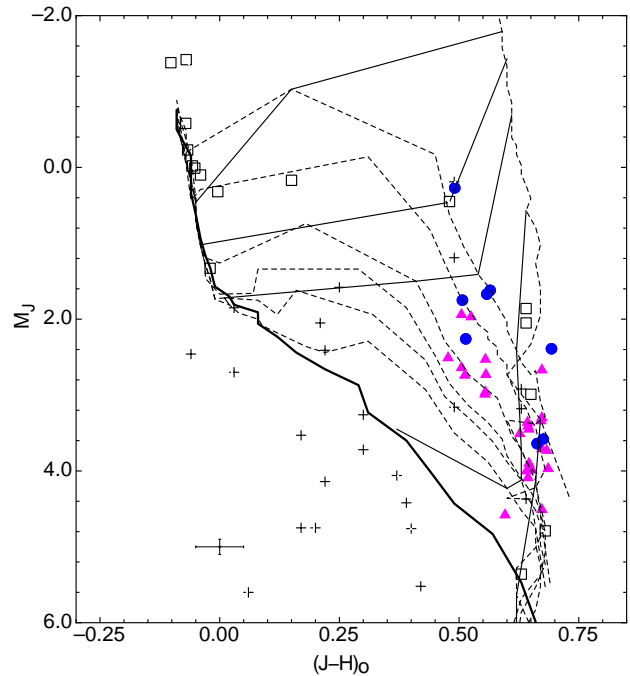}
\caption{Near-IR colour-magnitude diagram for stars in our sample. CTTs and WTTs are respectively represented by filled circles and triangles. Open squares show the PMS star candidates and crosses indicate the remaining unclassified stars. Isochrones with ages of 0.1, 1, 2, 5, 8, 10 and 15 Myr (dashed lines) and the ZAMS (thick line) from Siess et al. (2000). The thin lines represent the tracks for masses of 0.2, 1, 2, 3 and 4 M$_{\odot}$.}
\label{cmd}
\end{figure}

Using the {\it 2MASS} data corrected for reddening, in Fig. \ref{cmd} we show the {\it $M_J$ vs. (J-H)$_o$} colour-magnitude diagram with isochrones ranging from 0.1 Myr to the zero-age main sequence (ZAMS) from \cite{Siess2000}. Adopting ranges defined by the isochrones separation, we can evaluate the distribution of ages in our sample. The majority of the identified young stars are mainly in the 0.1 - 5 Myr range, in agreement with the results obtained from measurements of the lithium line (Sec. 4.3) and also with the ages 
 $<$ 5 Myr determined for the cluster close to Z CMa \citep{GregorioHetem2009}  and 1.5 Myr for a cluster associated with the BRC~27 cloud (Soares et al. 2002). Only 16\%  are in the range from 5 - 10 Myr and 1 object has an age above 10 Myr.

\begin{figure}
\centering
\includegraphics[width=8cm,angle=0]{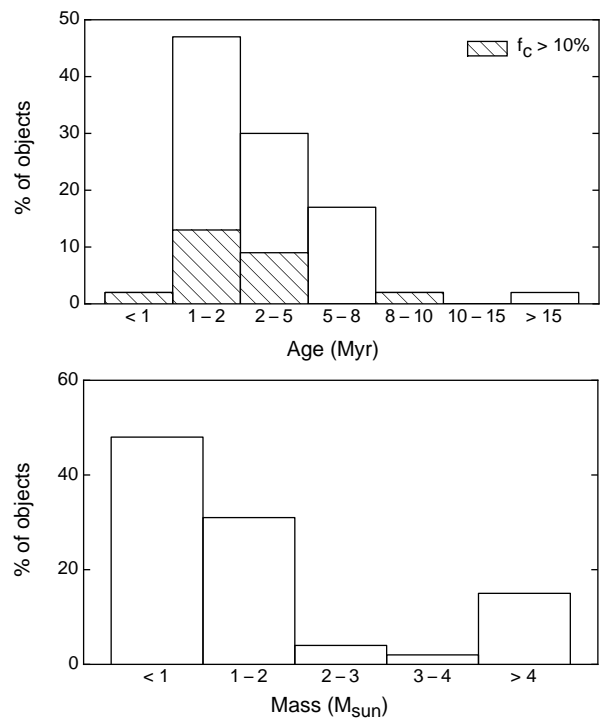}
\caption{Age and Mass distribution of the PMS stars and candidates. Hatched areas are used in top panel
to show the fraction of disc-bearing stars.} 
\label{histograms}
\end{figure}

Our results on age distribution are similar to those found by \cite{Oliveira2013b} for the members of the Serpens cloud. The comparison between the age distribution of the discless and disc-bearing stars coexisting in Sh~2-296 show no significant difference. Among the 12 objects showing $f_c > 10$\% and having an estimate of age, six (No. 5a, 13, 22, 23, 30 e 34a) are in the range 1-2 Myr, four (No. 10b, 14, 47, 55) are in the range 2-5 Myr, one (No. 18) is $<$ 1Myr and another one (No. 9) is 8-10 Myr.  
A comparison of this distribution with the histogram of ages of our sample (see Fig. \ref{histograms} top panel) suggests
that stars without discs are not typically older, but rather had shorter disc lifetimes.

The near-IR colour-magnitude diagram also shows that  48\% of the stars has less than 1 M$_{\odot}$ and 31\% 
 is in the range of 1 - 2 M$_{\odot}$, consistent with the expected masses for T Tauri stars. Only five candidates have more than 4 M$_{\odot}$, which were previously identified as B-type stars in the literature. The distribution of masses and ages is represented by the histograms in Fig. \ref{histograms}.

Among the additional stars, 52\% is located below the ZAMS, meaning that they are most likely in a different distance from Sh~2-296 and probably are not associated with the cloud. Ages and masses were
estimated for 11 additional stars, which possible classification is discussed in the Appendix A.

\subsection{Confirming the nature of the candidates}

The detailed analysis of the photometric characteristics and SED fitting allowed us to better evaluate the nature of candidates that could not be unambiguously identified as young stars by our spectroscopic characterization. In this section we discuss the reasons for their classification, which is giving in Table \ref{ysos}.

\ 

\noindent {\it T Tauri stars}

\ 

Object 07035805--1132398 (No. 25) is located in the interface between WTT and CTTs according both the criteria presented in Sec. 4.2. However, the  lack of near-IR excess suggest this object being most likely a WTT.

The candidates 07035542--1135149 (No. 18) and 07035043--1133425 (No. 9) have infrared colours consistent with T Tauri stars, late spectral type and show very strong H$\alpha$ emission. The non-detection of the Li I line is most likely caused by veiling. According to the criteria presented in Sec. 4.2, these objects would be classified as CTTs.

Object  07040224--1124188 (No. 33) also did not have the Li line detected and it shows H$\alpha$ emission combined with absorption, suggesting variability of the line. It is located in the region F at the near-infrared colour-colour diagram and could be either a reddened main-sequence star or a young object with little or no near-IR excess. However, its position compared to the isochrones on the colour-magnitude diagram, suggests that this object has an age of about 1 Myr and is probably a T Tauri star.

For object 07042625-1131207 (No. 60) we acquired a very low quality spectrum and could not determine spectral type or measure the Li line, but it has a strong H$\alpha$ emission. This object is a X-ray emitting source and also shows significant infrared excess, more distinguishably at  4.6 $\mu$m K-[4.6] = 0.87). Assuming an average visual extinction of  Av = 1.0 mag, the location at the colour-magnitude diagram suggests an age of less than 1 Myr and a mass lower than 2 M$_\odot$, thus we classify this object as a 
possible CTT.

\ 

\noindent {\it Intermediate-mass stars}

\ 

The near-IR colour-magnitude diagram in Fig. \ref{cmd} clearly shows a group of intermediate-mass objects ($>$3 M$_{\odot}$) near to the ZAMS, which were classified with the generic term ``PMS"
since their spectrum did not show H$\alpha$ in emission.

Objects 07035152-1134557 (No. 10a), 07035486-1134340 (No. 16) 07035880-1135311 (No. 27), 07040234-1125393 (No. 34b), 07040393-1126097 (No. 40), 07040542-1128562 (No. 43) and 07041588-1124055 (No. 45) are associated with X-ray  emission similar to those of other PMS stars in our sample. They have been identified as early type YSO candidates associated with CMa R1 by  Herbst, Racine and Warner (1978) and \cite{Shevchenko1999}.  

\cite{Rebull2013} also studied two of these objects (No. 34b and 40) as part of the bluest YSO candidates in the region of BRC~27, however their analysis of the SEDs could not confirm the young nature of these objects since they did not detect IR excess for No. 40 and only identified an uncertain IR excess for No. 34b. We were only able to evaluate the SED for No. 40, since No. 34b has no {\it WISE} data, and have also not found any contribution from circumstellar emission.

Object  07035564-1132468 (No. 19) presents X-ray emission consistent with other identified PMS candidates, however we have not observed H$\alpha$ emission or infrared excess. The SED fitting derived a temperature of 8000 K for this star and its position on the colour-colour and colour-magnitude diagrams suggests  this is a PMS object not identified in the previous works found in the literature.

Another intermediate mass object is 07040225--1125429 (No. 34a). We detected H$\alpha$ emission in its spectrum and it has also been previously identified as an emission line star associated with BRC~27 by \cite{Ogura2002} and mentioned by \cite{Rebull2013} as needing confirmation as a young star. The inspection of the SED, near- and mid-IR colour-colour diagrams in Fig. \ref{diagrams}  shows that this object has significant IR excess possibly associated with a circumstellar component and an age between 1 and 5 Myr. We classify this object as a Herbig Ae candidate.

\ 

\noindent {\it Probable field stars}

\

There are two other objects showing H$\alpha$ emission for which we have not detected the Li line: 07034606--1135120 (No. 3) and 07034651-1135118 (No. 4). For these objects we could not find any other confirmation of youth trough the photometric characteristics or SED fitting. For this reason, we consider they possibly are active  M-type field stars.

\subsection{Spatial Distribution}

\begin{figure*}
\includegraphics[angle=270,width=0.953\textwidth]{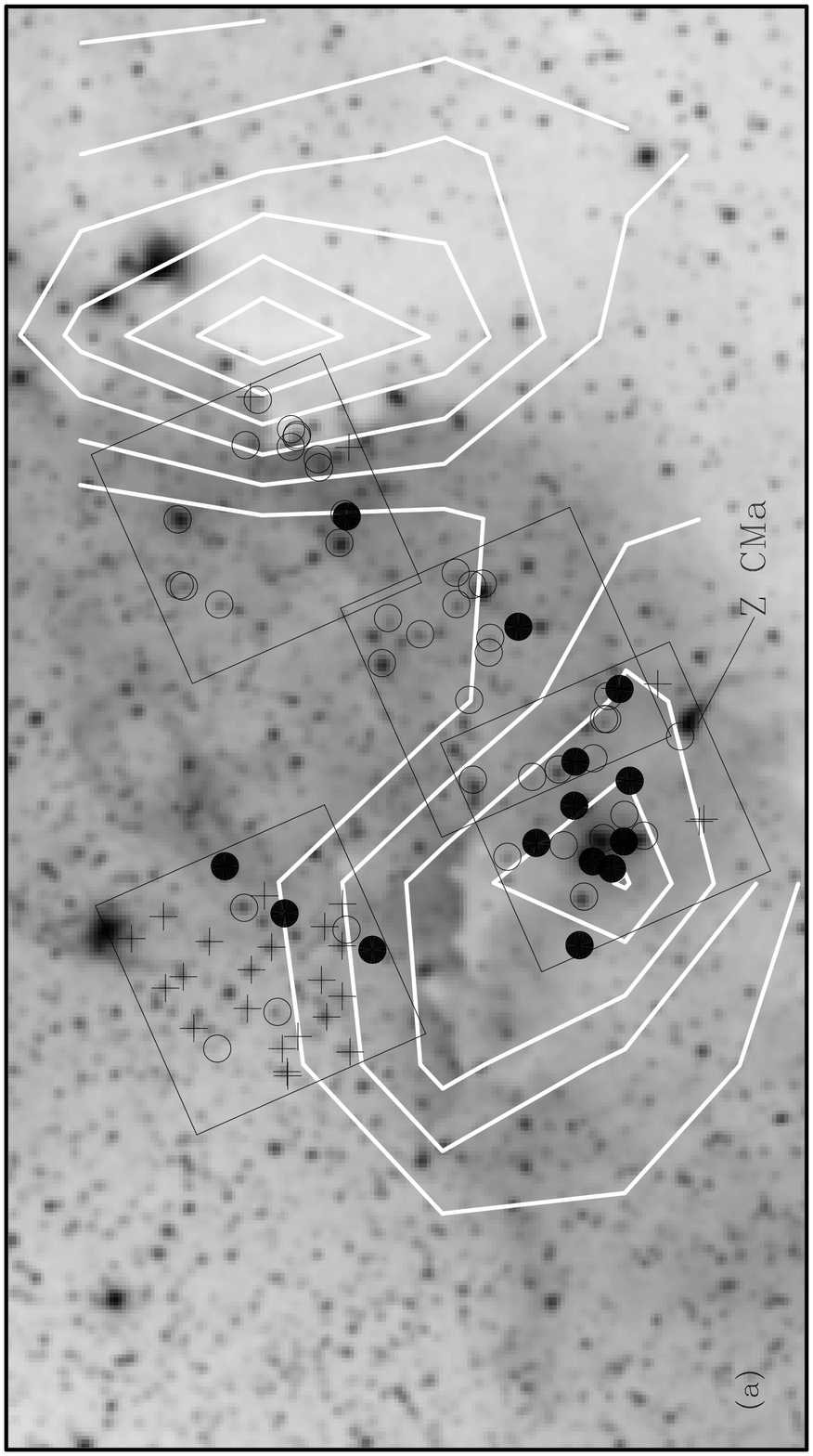}\\
\includegraphics[angle=270,width=0.953\textwidth]{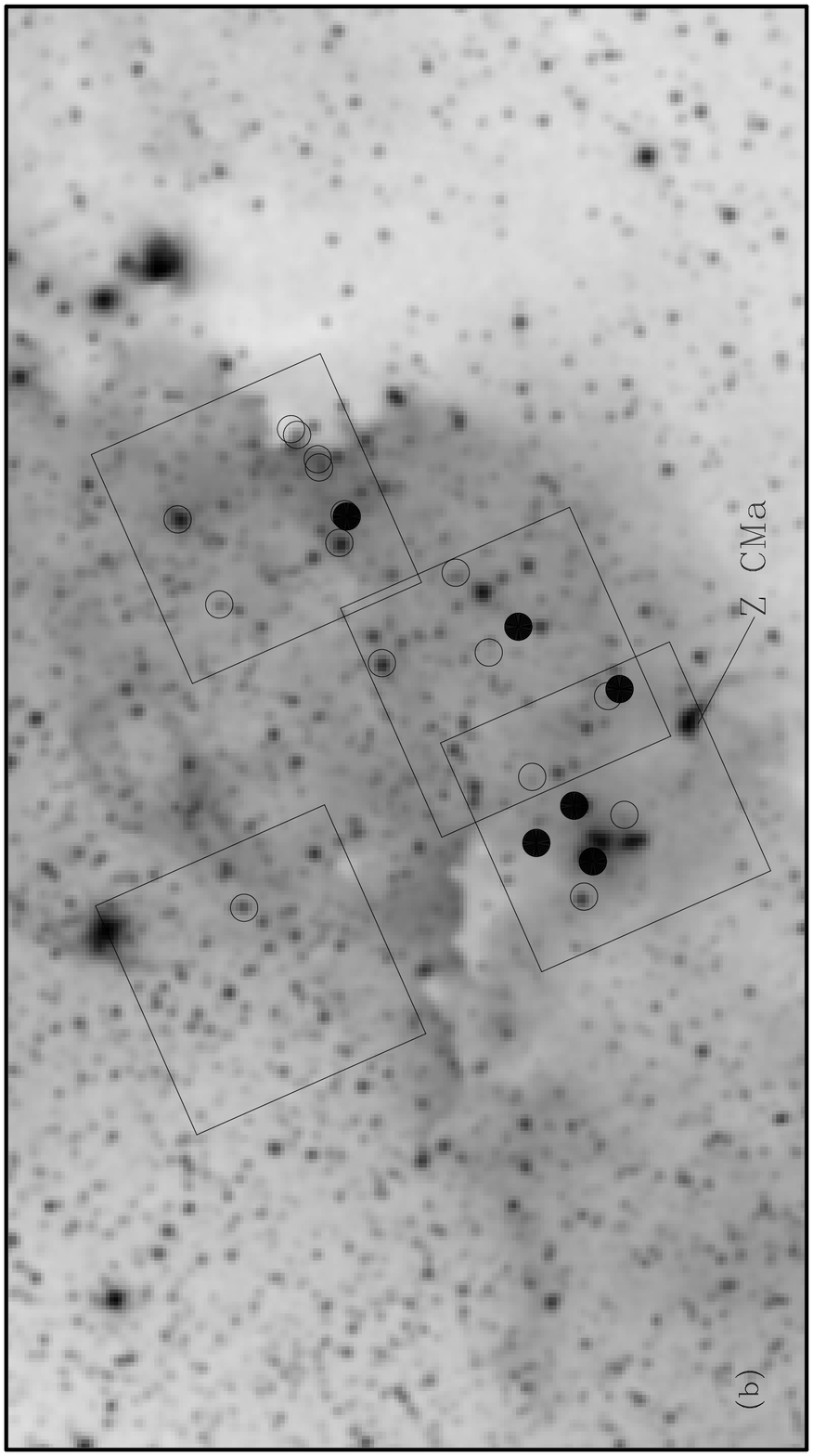}
\caption{Optical DSS image of Sh~2-296 showing  the fields observed with GMOS, which is a zoom of Fig. \ref{fieldtot} top panel. The PMS stars are represented by circles. Filled circles show the disc-bearing objects, which have $f_c >$ 10\%.  (a): Contours of $^{12}$CO emission in white (Mizuno \& Fukui 2004). Un-classified stars are represented by crosses. (b): Distribution of stars with estimated ages in the range of 1-2 Myrs.}
\label{dist_tts}
\end{figure*}

Figure \ref{dist_tts}a shows the distribution of the objects identified as young stars in our sample. The GMOS field CMaX~66, encompassing the star Z~CMa  and the cluster VDB~92, contains 40\% of the young stars revealed in the present work. The other two studied fields (CMaX~71 and CMaX~75) that also coincide with the edge of Sh~2-296 present, each one, 24\% of the detected young stars. On the other hand, the field CMaX~89 that is located to the east side of the Sh~2-296 edge, contains only 12\% of the young stars. 

While almost half of the young stars have an age of 1-2 Myrs or less (with a mass of less than 1-2 $M_\odot$), comparable to that of the cluster around Z CMa \citep{GregorioHetem2009}, only a small fraction ($\sim 25\%$) show evidence of IR excess revealing the presence of circumstellar discs. While examples of such a low fraction exist (e.g., Hernandez et al. 2008), they are quite rare compared to most young star-forming regions (for instance, the $\rho~Oph$ region shows almost 100$\%$; see, e.g., review by Hillenbrand 2005). This suggests that some external factor has accelerated the disc dissipation. Possible mechanisms include stellar winds from nearby O stars, like in the Sco-Cen association, or intense UV radiation, either from the star itself (if sufficiently massive), or, again, from nearby O stars as in Orion -although this process generally requires several million years \citep{Allen2012}.

However, when one considers the overall spatial distribution of these young stars (and more particularly those in the 1-2 Myr age interval, shown on Fig. \ref{dist_tts}b), an additional, important property appears: there is a strong segregation between the disc-bearing stars (CTTs), clearly tied to the two dense molecular clouds in the region, and the discless stars, which are all seen inside the nebula (at least in projection). This segregation is reinforced when one considers the rest of the {\sl XMM-Newton} sources, because, except for one case (J07045632-1129332), none of them show IR excess from 2MASS. These X-ray emitting stars are not characterized spectrally like those in the {\sl Gemini} fields, but we have checked that they have $L_X/L_{bol} \approx 10^{-4}-10^{-3}$), so must be also young stars (see Santos Silva et al., in preparation).

A similar distribution of PMS stars is found, for example, in the Lupus star-forming region, consisting of four molecular clouds. \cite{Galli2013} evaluated the location of young members of a moving-group, finding a result similar to ours: the stars identified as CTTs are associated with higher levels of $^{12}$CO, which traces the molecular clouds, while only WTTs are found outside the clouds. However, the authors find that the WTTs in Lupus tend to be older than the CTTs, a trend we do not detect in the present work.

\section{Summary and concluding remarks}

We used optical spectroscopy to select and characterize counterparts of a sample of X-ray sources detected in the Sh~2-296 nebula, associated with the Canis Major R1 star-forming region. The spectroscopic data was complemented with photometry from different publicly available catalogues, covering a wide wavelength range from the optical (0.44 $\mu$m) to the mid-IR (22 $\mu$m).

Among 83 stars observed with {\it Gemini+GMOS} we find 41 showing Li line that confirms them as T Tauri stars. The remaining 42 objects were separated in two groups: 17 are considered young star candidates, since they show some of the typical characteristics (H$\alpha$ emission, X-ray source and/or association with Sh~2-292 previously reported in the literature); and 25 are additional objects that do not coincide with X-ray sources and have  H$\alpha$ in absorption. A tentative classification for the group of additional stars is discussed in Appendix A.

We have determined spectral types for the majority of the sample, by comparing the continuum and TiO band shapes to libraries of stellar spectra. Using criteria with basis in the H$\alpha$ emission we were able to distinguish between Classical and Weak T Tauri in our sample, finding a CTT fraction of 17\%. 

The nature of the candidates was also investigated by using near- and mid-IR photometry as indicator of the circumstellar characteristics. Colour-colour and colour-magnitude diagrams revealed the objects showing IR excess and allowed us to estimate ranges of mass and age.  The fraction of circumstellar emission was estimated by a SED fitting, performed with the code GADisk. 

The IR characterization indicates that 5 young star candidates showing H$\alpha$ emission and/or association with X-ray source  probably are T Tauri stars, 4 of them possibly are CTTs. If this suggestion is confirmed, the total number of T Tauri stars in our sample would increase from 41 to 46, while the number of CTTs would change from 7 to 11. In this case, the fraction of CTTs in our sample would be of 24\%. 

The position near to the ZAMS in the colour-magnitude diagram indicate that 9 young star candidates not showing H$\alpha$ emission, neither IR excess are {\it indeed} PMS objects, confirming the identification found in the literature as B stars associated with Sh~2-296 for 8 of them.

Among the objects indicated as YSO candidates in the literature, the stars No. 5a, 20a, 36a and 50 are confirmed in the present work as T Tauri stars. We were able to identify No. 37, and 60 as probable T Tauri stars, by the analysis of their photometric characteristics. We also confirm the nature of object No. 34a as a Herbig Ae star, which was previously identified by \cite{Rebull2013} but  still required spectroscopic confirmation of youth. These confirmations reinforce that to efficiently identify and characterize young stellar objects it is essential to evaluate data at multiple wavelengths. 

In summary, the suggested classification of the 17 candidates is: 5 T Tauri; 1 HAe; 9 PMS; 2 field stars. Among with the 41 confirmed T Tauri stars, in overall we revealed the youth of 56 stars in the region of Sh~2-296, 75\% of them not previously identified in the literature.

In spite of our investigation being restricted to a small area, a trend of spatial distribution can be noted. Altogether there is no disc-bearing young population within the nebula. This tends to suggest that the nebula itself is somehow responsible for the early disappearance of the discs for these stars. In the absence of relevant data, one can only speculate that either the stars have drifted into a static nebula (velocities as low as $\sim 1 {\rm km~s}^{-1}$ are sufficient), or the nebula has expanded quickly to overtake the recently formed stars. However, the implied disc dissipation time of 1-2 Myr is too short to be explained by UV radiation (see, e.g., the discussion in \cite{Allen2012}), leaving open the possibility of mechanical (aerodynamic) disc dissipation due to passage of a shock wave, i.e., resulting from the supernova explosion of a nearby massive star. In general, supernova remnants leave a clear signature in the form of diffuse X-ray emission, but this is excluded by our own {\sl XMM-Newton} observations. However, the supernova remnant may have cooled in less than 1 Myr and disappeared if the medium in which it had expanded was dense enough (as in the outer regions of a molecular cloud, 100 cm$^{-3}$ or more) such as the ones we now see in CO; the present HII region could then be ``fossil'' in the sense that it would be the (short-lived) end result of this cooling.

In the end, the star formation history of the young cluster associated with Z CMa appears strongly influenced by the presence of the Sh~2-296 nebula. However, because of the absence of nearby OB stars like in Sco-Cen, Cepheus or Orion, we are left with a new puzzle: not only can we only speculate about the source of the ionizing photons of Sh~2-296 itself, but it seems difficult to explain the early disappearance of the discs of the stars inside the nebula, which may -or may not- be related to it. We can now only conclude that the Sh~2-296 nebula must have originated at least 1-2 Myr ago, which is much longer than the time scale for recombination of the HII region. Perhaps a past supernova remnant, having cooled by now, remains the only viable explanation.

A more extensive and detailed study of the young stellar population in this region is still required to help elucidate the star formation history of CMa R1 and the origins of Sh~2-296. Our team is developing an ongoing work that expands our sample to about 400 new X-ray sources discovered by us (Santos-Silva et al. in preparation). We are already performing analysis and acquiring spectroscopic data for these new candidates.

\bigskip

\section*{Acknowledgments}

The authors are grateful to the anonymous referee for the important suggestions/corrections on this paper. Part of this work was supported by CAPES/Cofecub Project 712/2011. BF thanks CNPq Project 142849/2010-3. JGH acknowledges partial support from FAPESP Proc. No. 2010/50930-6.
This publication makes use of data products from the Two Micron All Sky Survey, which is a joint project of the University of Massachusetts and the Infrared Processing and Analysis Center/California Institute of Technology, funded by the National Aeronautics and Space Administration and the National Science Foundation.
This publication makes use of data products from the Wide-field Infrared Survey Explorer, which is a joint project of the University of California, Los Angeles, and the Jet Propulsion Laboratory/California Institute of Technology, funded by the National Aeronautics and Space Administration.

\appendix
\section{Additional objects}

Following the criteria described in Sect. 4, among our sample of 83 stars observed with {\it GMOS} we selected 58 possible PMS stars, which in the first analysis were considered as T Tauri (41) or candidates (17). The other 25 stars having none of the spectral features used to select the PMS candidates were
considered additional objects. Since the signal-to-noise ratio is good for 85\% of our sample, 
the lack of H$\alpha$ emission or Li line suggest that these 25 undefined objects possibly are not PMS,
 mainly because all of them are not related to X-ray emission. 
 
However, the resolution of the {\it GMOS} spectra could not be sensitive enough to detect lithium line 
with W(Li)$<$ 0.2\AA, which is expected for F-G stars with ages $>$ 20 Myr, for instance. These older stars
could also be associated to Sh~2-296, as part of an earlier star formation event. In order to be more
conservative, the 25 additional stars were also characterized on the basis of their near-IR colours (when
available), similar to the method adopted in Sect. 5.1 and Sect. 6.1. 

The results for these additional stars are presented in Tab. \ref{a1}. Spectral types were estimated by comparing the observed with the library spectra. For stars earlier than K0 we used the  6000-6700 \AA \ region, due to the lack of  TiO bands for these stars. For these cases we were only able to estimate spectral types for objects with S/N $>$ 15 and the visual comparison seems to result in an uncertainty of up to $\sim$ 5 spectral types.

Infrared colours were derived from 2MASS data that are available with good quality for 92\% of this 
sub-sample. The colours shown in Tab. \ref{a1} were corrected for
extinction by adopting a mean value of A$_V$=1.0 mag. The position of these additional objects in Fig. \ref{diagrams}a (black crosses) is consistent with Main-Sequence stars, except for objects 5b and 58. These two stars show high K excess (H-K $>$ 0.4) that could indicate they are PMS candidates. However, their absolute magnitude $M_J >$ 5 indicate they are not at the same distance of Sh~2-296. 

The 14 objects appearing below the ZAMS in Fig. \ref{cmd}
were considered unidentified stars, indicated by ``??" mark in Tab. \ref{a1}. The other 11 objects that are located inside the range of the isochrones in the colour-magnitude diagram could have an estimation of ages and masses. We consider these stars as possible young stars, classified as ``PMS?", 45\% of them  are $<$ 5 Myr, 10\% is 5-10 Myr, 27\% is 10-20 Myr and 18\% is $>$ 20 Myr.
Masses range from 1 to 3 M$_{\odot}$.  

\begin{table}
\caption{ Additional objects observed with {{\it GMOS}}.}
\begin{center}
{\scriptsize
 \begin{tabular}{cccccccc}
  \hline
ID	&	S.T	&	(J-H)$_o$	&	M$_J$	&	age	&	mass	&	Class.	&	comment	\\
 \hline
2	&		&		&		&		&		&	??	&	no JHK/low S/N	\\
5b	&		&	0.42	&	5.52	&		&		&	??	&	IR excess/low S/N	\\
28	&	F4	&	0.40	&	4.76	&		&		&	??	&		\\
42	&		&	0.64	&	4.37	&	10 - 15	&	0.2 - 1	&	PMS?	&	low S/N	\\
49	&	G0	&	0.39	&	4.42	&		&		&	??	&		\\
51	&	G7	&	0.25	&	1.58	&	$\sim$ 3	&	8 - 10	&	PMS?	&		\\
52	&	F6	&	0.30	&	3.72	&		&		&	??	&		\\
53	&	F4	&	0.22	&	4.14	&		&		&	??	&		\\
54	&		&	0.63	&	2.92	&	1 - 2	&	$\sim$ 1	&	PMS?	&	low S/N	\\
56	&		&	-0.06	&	2.46	&		&		&	??	&	low S/N	\\
57	&		&	0.49	&	0.19	&	$<$ 1	&	0.1 - 1	&	PMS?	&	low S/N	\\
58	&	G0	&	0.06	&	5.60	&		&		&	??	&	IR excess	\\
59	&		&	0.03	&	2.70	&		&		&	??	&	low S/N	\\
62	&	F4	&	0.20	&	4.75	&		&		&	??	&		\\
64	&	K4	&	0.49	&	3.16	&	$\sim$ 10	&	1 - 2	&	PMS?	&		\\
65	&	K2	&	0.63	&	3.18	&	1 - 2	&	$\sim$ 1	&	PMS?	&		\\
66	&	G4	&	0.22	&	2.41	&	$\sim$ 15	&	1 - 2	&	PMS?	&		\\
67	&	G8	&		&		&		&		&	??	&	bad JHK	\\
68	&	G7	&	0.30	&	3.26	&	ZAMS	&	1 - 2	&	PMS?	&		\\
69	&	F4	&	0.37	&	4.06	&		&		&	??	&		\\
70	&	F5	&	0.17	&	4.75	&		&		&	??	&		\\
71	&		&	0.03	&	1.85	&	ZAMS	&	$\sim$ 3	&	PMS?	&	low S/N	\\
73	&		&	0.21	&	2.05	&	10 - 15	&	1 - 2	&	PMS?	&	low S/N	\\
74	&	K2	&	0.49	&	1.19	&	1 - 2	&	2 - 3	&	PMS?	&		\\
75	&	F9	&	0.17	&	3.53	&		&		&	??	&		\\
\hline
\label{a1}
\end{tabular}
}
\end{center}
\end{table}

\end{document}